\documentclass[12pt]{article}
\usepackage{}
\usepackage{epsfig}

\usepackage{a4}
\usepackage{amsmath}

\usepackage{epsfig}
\usepackage{amsmath}
\usepackage{amsfonts} 
\usepackage{amssymb}
\usepackage{ifthen}

\begin{document}
\title{Vortices of $SO(2)$ gauged Skyrmions in $2+1$ dimensions}
\author{{\large Francisco Navarro-L\'erida}$^{1}$
and {\large D. H. Tchrakian}$^{2,3}$ 
\\ 
\\
$^{1}${\small Dept.de F\'isica Te\'orica, Ciencias F\'isicas,}\\
{\small Universidad Complutense de Madrid, E-28040 Madrid, Spain}\\  
$^{2}${\small School of Theoretical Physics, Dublin Institute for Advanced Studies,}
\\{\small 10 Burlington Road, Dublin 4, Ireland }
\\
  $^{3}${\small  Department of Computer Science, NUI Maynooth, Maynooth, Ireland}}

\date{\today}
\newcommand{\dd}{\mbox{d}}
\newcommand{\tr}{\mbox{tr}}
\newcommand{\la}{\lambda}
\newcommand{\om}{\omega}
\newcommand{\ka}{\kappa}
\newcommand{\ta}{\theta}
\newcommand{\f}{\phi}
\newcommand{\vf}{\varphi}
\newcommand{\vr}{\varrho}
\newcommand{\F}{\Phi}
\newcommand{\al}{\alpha}
\newcommand{\bt}{\beta}
\newcommand{\ga}{\gamma}
\newcommand{\de}{\delta}
\newcommand{\si}{\sigma}
\newcommand{\Si}{\Sigma}
\newcommand{\bomega}{\mbox{\boldmath $\omega$}}
\newcommand{\bsi}{\mbox{\boldmath $\sigma$}}
\newcommand{\bchi}{\mbox{\boldmath $\chi$}}
\newcommand{\bal}{\mbox{\boldmath $\alpha$}}
\newcommand{\bpsi}{\mbox{\boldmath $\psi$}}
\newcommand{\brho}{\mbox{\boldmath $\varrho$}}
\newcommand{\beps}{\mbox{\boldmath $\varepsilon$}}
\newcommand{\bxi}{\mbox{\boldmath $\xi$}}
\newcommand{\bbeta}{\mbox{\boldmath $\beta$}}
\newcommand{\ee}{\end{equation}}
\newcommand{\eea}{\end{eqnarray}}
\newcommand{\be}{\begin{equation}}
\newcommand{\bea}{\begin{eqnarray}}

\newcommand{\ii}{\mbox{i}}
\newcommand{\e}{\mbox{e}}
\newcommand{\pa}{\partial}
\newcommand{\Om}{\Omega}
\newcommand{\vep}{\varepsilon}
\newcommand{\bfph}{{\bf \phi}}
\def\theequation{\arabic{equation}}
\renewcommand{\thefootnote}{\fnsymbol{footnote}}
\newcommand{\re}[1]{(\ref{#1})}

\newcommand{\eins}{1\hspace{-0.56ex}{\rm I}}
\newcommand{\R}{\mathbb R}
\newcommand{\C}{\mathbb C}
\newcommand{\p}{\mathbb P}
\renewcommand{\thefootnote}{\arabic{footnote}}

\maketitle


\bigskip

\begin{abstract}
  Vortices the $SO(2)$ gauged planar Skyrme model , with {\bf a)} only Maxwell, {\bf b)} only Chern-Simons, and {\bf c)} both
  Maxwell and Chern-Simons dynamics are studied systematically. In cases {\bf a)} and {\bf b)}, where both models feature a single
  parameter $\la$ (the coupling of the potential term), the dependence of the energy on $\la$ is analysed. It is shown that the
  plots of the energy $vs.$ $\la$ feature discontinuities and branches. In case {\bf c)}, the emphasis is on the evolution of the
  topological charge, taking non-integer values. Throughout, the properties studied are contrasted with those of the corresponding Abelian Higgs models.

\end{abstract}
\medskip
\medskip

\section{Introduction}
Solitons of gauged Skyrme model, Skyrmions,
present much more complex properties than their Higgs counterparts, $e.g.$ the Abelian Higgs vortices
in $2+1$ dimensions~\cite{Nielsen:1973cs}, monopoles in $3+1$ dimensions~\cite{'tHooft:1974qc,Polyakov:1974ek} and monopoles in
$D+1$ dimensions~\cite{Tchrakian:2010ar}. One reason for this is that the topological
charge densities of gauged Higgs systems can be
systematically constructed with a $top\ down$ approach~\cite{Tchrakian:2010ar} by subjecting
Chern-Pontryagin densities to dimensional descent, while the topological charge densities of gauged Skyrme
models, as defined in Ref.~\cite{Tchrakian:1997sj}, can be constructed instead {\bf only} with a $bottom\ up$
approach~\footnote{Indeed this $bottom\ up$ approach can be applied also to Higgs field systems~\cite{Tchrakian:2002ti}.
In that case, the $bottom\ up$ definition coincides with the $top\ down$ definition
of the topological charge only for the Higgs models on $\R^2$.}.

By gauged Skyrmion we understand the soliton of a $SO(N)$ gauged $O(D+1)$ sigma model~\footnote{The most general Lagrangian
of a Skyrme model, whether gauged or ungauged, consists of a potential term plus quadratic, quartic, and up to $2D$-atic
kinetic terms. This is to satisfy the Derrick scaling requirement.}
in $D+1$ dimensional spacetime, or on $\R^D$, with $D\ge N\ge 2$, supporting topologically
stable solitons. In the case at hand, this is the $SO(2)$ gauged $O(3)$ sigma model.
It is understood that the vacuum of the gauged model is the same as the vacuum of the gauge-decoupled model, such that
the gauge symmetry is not broken in the vacuum. This is in stark contrast to (all) Higgs models, whose dynamics explicitly
displays (gauge) symmetry breaking.

Our purpose is to investigate the detailed properties of $SO(2)$ gauged Skyrmions whose dynamics is controlled by a Maxwell (M) term, or a Chern-Simons (CS) term, or both (MCS). The influence of the CS dynamics in the model incorporating both Maxwell and Chern-Simons terms, was studied recently~\footnote{Similar models were used in Refs.~\cite{Samoilenka:2016wys,Samoilenka:2015bsf}, with various self-interaction potentials, where the solutions were constructed subject to no symmetries, where the effect of the Chern-Simons term on the interaction of the solitons was investigated.} in (Section {\bf IV} of)
Ref.~\cite{Navarro-Lerida:2016omj}. There the emphasis was laid on the (unusual) dependence of the energy on the electric charge
and the angular momentum. Here by contrast, our attention will be focused on the evolution of the topological charge, which we have recently considered both for $2+1$ and $3+1$ dimensional $SO(2)$ gauged Skyrmions, in \cite{Navarro-Lerida:2018giv}.

Topologically stable vortices of the $SO(2)$ gauged planar Skyrmion with Maxwell
dynamics were first given by Schroers~\cite{Schroers:1995he} and later in Ref.~\cite{Gladikowski:1995sc}. The vortices constructed in \cite{Schroers:1995he} were self-dual, while the vortices presented in \cite{Gladikowski:1995sc}~\footnote{An interesting question
studied in \cite{Gladikowski:1995sc} was the dependence of the topological lower bound on the gauge coupling constant.}
were solutions of second-order equations for a model which in the gauge-decoupling limit reduced to the (ungauged) planar Skyrme model.
The definition of the topological charge proposed in \cite{Schroers:1995he} was subsequently extended to
higher dimensions in Ref.~\cite{Tchrakian:1997sj}. (See also
Appendix {\bf B} of \cite{Tchrakian:2015pka}, and references therein.) Employing the Chern-Simons term as
an alternative for the Maxwell dynamics was considered in Refs.~\cite{Ghosh:1995ze,Kimm:1995mi,Arthur:1996uu},
in analogy with its Higgs counterpart studied in Refs.~\cite{Hong:1990yh,Jackiw:1990aw}.

In the present work we consider {\bf a)} the model whose
gauge dynamics is described by the Maxwell term, namely the model studied in
Ref.~\cite{Schroers:1995he}, {\bf b)} the model whose
gauge dynamics is described by the Chern-Simons term, studied in Refs.~\cite{Ghosh:1995ze,Kimm:1995mi,Arthur:1996uu}, and
{\bf c)}, the model whose gauge dynamics is described both by the Maxwell and the Chern-Simons terms, $e.g.$ as in
\cite{Navarro-Lerida:2016omj}, where the emphasis was different.


We have two distinct but not unrelated objectives. The {\it first} objective is to reveal the peculiar dependence of the energy on a parameter in the model,
that is typical of gauged Skyrmions. This has been studied in the case of the $SO(3)$ gauged Skyrmion on $\R^3$, namely the $O(4)$ sigma model, where the energy was plotted
against the coupling of the (quartic kinetic) Skyrme term. It was found in \cite{Brihaye:1998vr,Brihaye} and in \cite{Kleihaus:1999ea} that the energy profile exhibited {\it discontinuities} and {\it branches}. This contrasts with the corresponding Higgs model, namely the $SO(3)$ gauged Georgi-Glashow model supporting the monopole~\cite{'tHooft:1974qc,Polyakov:1974ek}, where the energy profile plotted against the coupling strength of the Higgs self-interaction potential is shown~\cite{Kirkman:1981ck} to be {\it continuous} and {\it monotonic} and as it happens, flattens out for large Higgs coupling. The situation in the case of the $SO(2)$ gauged Higgs model on $\R^2$, namely the Abelian Higgs model supporting vortices, is qualitatively similar to the three dimensional case. It is shown in Ref.~\cite{Burzlaff:2010hh} that  energy profiles plotted against the Higgs coupling are {\it continuous} and {\it monotonic}, and increase logarithmically for large Higgs coupling. The corresponding study for any $SO(2)$ gauged Skyrmion in $2+1$ dimensions, $either$ with Maxwell~\cite{Schroers:1995he}, $or$ with Chern-Simons~\cite{Ghosh:1995ze,Kimm:1995mi,Arthur:1996uu} dynamics, has not been carried out. This is our first objective. Our analysis here will confirm that also in the two dimensional case, some energy profiles exhibit discontinuities and branches.
                    
Our $second$ objective, which from the viewpoint of dynamics is the dominant one, is to track the evolution of the topological charge in a given gauged Skyrme model. (The relationship of the energy with the electric charge and the spin are already studied in Ref.~\cite{Navarro-Lerida:2016omj}.) The gauge-decoupled
Skyrmion has integer ``baryon number'' which is calculated as the volume integral of the winding number density.
Introduction of the
$SO(2)$ gauge field alters the definition of the topological charge density, and further, in the presence of
Chern-Simons dynamics the
integral of the altered topological charge density departs from the integer value.
This mechanism is predicated on the presence
of the electric component of the gauge potential, which in turn results from the presence of the Chern-Simons term.
(Not surprisingly this mechanism is also responsible for the $unusual$ energy/electric charge and
energy/angular momentum dependence, which was the
subject of investigation in Ref.~\cite{Navarro-Lerida:2016omj}.)

The most general Lagrangian considered is
\be
\label{mostLag}
{\cal L}_{(\la)}=-\frac14\mu F_{\mu\nu}^2+\ka\vep^{\la\mu\nu}A_\la F_{\mu\nu}
-\frac18\tau D_{[\mu}\f^aD_{\nu]}\f^b+\frac12\eta^2|D_\mu\f^a|^2-\eta^4\,V[\f]\, .
\ee
                    
In achieving our {\it first} objective, we employ the two models {\bf a)} With Maxwell term only, $i.e.$, with $\ka=\tau=0$. {\bf b)} With Chern-Simons term only, $i.e.$, with $\mu=\tau=0$. We study the energy profiles plotted against the coupling strength of the potential term, with special emphasis to the first-order self-dual solutions in each case. The models considered do not contain quartic kinetic ``Skyrme'' terms since inclusion of these would prevent the saturation of the topological lower bound, and the study of $first\ order$ solutions is the main feature of these models.

To achieve our $second$ objective, we employ the model {\bf c)} incorporating both Maxwell and Chern-Simons terms, $i.e.$, with nonvanishing couplings $\mu$ and $\ka$ in \re{mostLag}.
This is the simplest system which enables the tracking of the topological charge leading to its annihilation. We have however incorporated the $quartic$ Skyrme kinetic term as well with coupling $\tau$.
The first reason is that the model in question should, in the gauge decoupling limit, support the ungauged Skyrmion characterised by the ``baryon number'' prior to gauging. This necessitates the
inclusion of the $quartic$ Skyrme term. It turns out also, that the presence of this term renders the numerical computations simpler. It is important to note that to achieve this aim, the potential
$V[\f]$ is chosen suitably as explained in Appendix {\bf C}.


Our study of {\bf a)} the gauged planar Skyrmion with Maxwell dynamics is presented in Section {\bf 2},
where we have employed two different potentials. The first one is the potential employed in Ref.~\cite{Schroers:1995he} which allows the construction of self-dual solutions.
The second model we study employs the ``pion mass'', where solutions to
the second-order equations are constructed. The numerical integrations here are simpler since the decay is exponential. In Section {\bf 3} we study {\bf b)} the
model with (only) Chern-Simons dynamics, where the new choice of potentials arises.  In both Sections {\bf 2} and {\bf 3}, we have studied the solutions of the second-order equations, in addition to those of the first-order equations. In Section {\bf 4} the combined Maxwell--Chern-Simons model {\bf c)} is studied, where there are only solutions to the second-order equations. Throughout, the said properties are studied
quantitatively using numerical methods. A summary of our results is given in Section {\bf 5}. The topological charge densities are defined in Appendix {\bf A},
and the resulting topological lower bounds on the energy densities (Belavin inequalities)
for Maxwell dynamics and for Chern-Simons dynamics are presented in Appendices {\bf B} and {\bf C} respectively. Since all the features revealed contrast starkly with the corresponding features in Higgs models, and since the crucial role is played by the topological charges and Belavin inequalities, we have inserted subsections in each Appendix exposing the (familiar) definitions for the corresponding Higgs models against which their Skyrme counterparts are contrasted.

Before proceeding to the study of the three types of models {\bf a)}, {\bf b)} and {\bf c)}, we state the results of symmetry imposition, including the topological and global charges,
which are common to all types of models {\bf a)}, {\bf b)} and {\bf c)}.

\subsection{Imposition of symmetry, topological and the global charges}
Since the field content of the models studied in the next three sections is the same one, we state the results of the imposition
of axial symmetry and the resulting charges here in advance. The definitions of the topological charges prior to the imposition of symmetry, are given in Appendix {\bf A}. For the definition of the familiar global charges, we refer to Ref.~\cite{Navarro-Lerida:2016omj}.

\subsubsection{Imposition of symmetry}
The planar Skyrme model, namely the $O(3)$ sigma model, is described by
the Skyrme scalar $\f^a=(\f^{\al},\f^3)$ subject to the constraint $|\f^a|^2=1$. The axially symmetric Ansatz is
\be
\label{axO3}
\f^{\al}=\sin f(r)\,n^{\al}\ ,\quad \f^3=\cos f(r)\ ,\quad
n^{\al}=\left(
\begin{array}{c}
\cos n\ta\\
\sin n\ta
\end{array}
\right)\,,
\ee
where $r$ is the radial coordinate, $\ta$ is the azimuthal angle and $n$ is the winding number of the Skyrme scalar.


Imposition of symmetry on the $SO(2)$ (Abelian) gauge field $A_\mu=(A_i,A_0)\ \mu=i,0$, is achieved by the Ansatz
\bea
A_i=\left(\frac{a(r)-n}{r}\right)(\vep\hat x)_i\quad&\Rightarrow&\quad F_{ij}=-\frac{a'}{r}\vep_{ij}\, , \label{Maxax}\\
A_0=b(r)\qquad\qquad\qquad\ \ &\Rightarrow&\quad\, F_{i0}=b'\, . \label{Maxax0}
\eea

The prescription of gauging employed is
\bea
D_\mu\f^{\al}&=&\pa_\mu\f^{\al}+A_\mu(\vep\f)^{\al}\ ,\quad(\vep\f)^{\al}=\vep^{\al\bt}\f^\bt \, , \label{coval}\\
D_\mu\f^3&=&\pa_\mu\f^3\label{cov3}\,,
\eea
and the covariant derivatives of $\f^a$ are
\bea
D_i\f^{\al}&=&(\sin f)'\,\hat x_in^{\al}-\frac{a\sin f}{r}(\vep\hat x)_i(\vep n)^{\al}\ ,\quad
D_0\f^\al=b\sin f\,(\vep n)^\al \, ,\label{Difal1}\\
D_i\f^{3}&=&(\cos f)'\,\hat x_i\ ,\qquad\qquad\qquad\qquad\qquad\ \ D_0\f^3=0\,.\label{Dif31}
\eea

Anticipating the usual choice of boundary conditions \re{skasymp} in the Appendix, the corresponding one-dimensional conditions read
\be
\label{bvf1}
\lim_{r \rightarrow 0} f(r) =\pi ,\qquad \lim_{r \rightarrow\infty} f(r) =0 \, ,
\ee
consistent with the finiteness of the energy, as well as with regularity at the origin. This choice of boundary values means that we are excluding the choice of symmetry-breaking
potentials~\footnote{Such potenials lead to a magnetic flux, and such models were considered in \cite{Loginov:2015jya}, investigating the relation of the magnetic flux and spin. Recently symmetry-breaking potentials in $2+1$ dimensions were employed
in \cite{Samoilenka:2017fwj}, where minimal energy multisolitons and their interactions were considered. This is not in the scope of the present work.} 

Concerning the boundary values of the function $a(r)$ in \re{Maxax} at the origin, regularity requires that
\be
\label{bva0}
\lim_{r \rightarrow 0} a(r) =n\,.
\ee
             As for $a(\infty)$, the boundary values of the function $a(r)$ at infinity, this will be left free. This is in stark contrast with usual~\footnote{The usual Abelian Higgs (AH) model is the $p=1$ AH model, for which $a(\infty)=0$ is forced by the requirement of finite energy. By contrast, all other AH models with $p \ge 2$ can support finite energy solutions~\cite{Navarro-Lerida:2016omj} for $a(\infty)\neq 0$.} Abelian Higgs model, where  $a(\infty)=0$.

\subsubsection{Topological charge}
Calculating the integral of $\vr_0$ defined by \re{3}, subject to \re{axO3}-\re{Maxax}, gives 
$\int\vr_0\,d^2x=-8\pi\,n\,.$
We thus denote the topological charge density of the ungauged Skyrmion as
\be
\label{baryon}
q_0=-\frac{1}{8\pi}\int\vr_0\,d^2x=n \, ,
\ee
which is the ``baryon number'' $n$.

After gauging, the topological charge is given by the density \re{9a}-\re{9b} and the resulting topological charge is
\be
\label{topchv}
q=-\frac{1}{8\pi}\left(\int\varrho_0\,d^2x+2\,\vep_{ij}\int \pa_{i} [(\phi^3-v)A_{j}]\,d^2x\right)\,.
\ee
When the choice $v=1$ is made, which as we have seen in Appendix {\bf B} is obligatory in the case when the Abelian
dynamics is controlled by the Maxwell term only, the topological charge equals the ``baryon number'' $n$.

With all other values of $v$, the topological charge departs from $n$.
As seen in Appendix {\bf C} the choice of $v\neq 1$ is viable in models featuring a Chern-Simons term, and the results obtained in the generic case are qualitatively the same. Here we have made the
choice $v=0$ throughout, as this is the simplest option.

With $v=0$ the topological charge is
\be
\label{topch}
q=-\frac{1}{8\pi}\left(\int\varrho_0\,d^2x+2\,\vep_{ij}\int \pa_{i} [(\phi^3)A_{j}]\,d^2x\right) \, ,
\ee
where the contribution of the second integral is not necessarily vanishing.

Evaluating \re{topch} subject to the symmetries \re{axO3}-\re{Maxax}, gives rise to
\be
\label{fintopch}
q=\frac12(n+a_\infty)\, ,
\ee
where $a_\infty = a(\infty)$.
The expression \re{fintopch} enables tracking the evolution of the topological charge (``baryon number'') by considering solutions with evolving values of $a_\infty$, $i.e.$, where $a_\infty\neq 0$. It is important here to stress that this is a strictly Skyrme theoretic mechanism, and is excluded by Higgs theoretic dynamics. Notably, $a_\infty=0$ for the usual ($p=1$) Abelian Higgs (AH) model~\footnote{It is however the case that vortices of the $p\ge 2$ AH models, alluded to in footnote $3$ above, with $a_\infty\neq 0$ exist. This fact does not lead to the evolution of the topological charge (the magnetic vortex number $\mu$). To see this we recall the expression for the topological charge of the $p$-AH model, which is descended~\cite{Tchrakian:2010ar} from the $2p$-th Chern-Pontryagin density. In terms of the radial functions $a(r)$ and $h(r)$ parametrising the radially symmetric $U(1)$ field and the Higgs field $\f^\al=h\,n^\al$, in the notation of Section {\bf 1.1.1} above, the expression for the topological charge corresponding to \re{fintopch} is
             \be
             \label{magtopch}
             \mu=\frac{1}{2\pi}\int\frac{d}{dr}[(1-h^2)^{2p-1}\,a]\,dr=[(1-h^2)^{2p-1}\,a]_{r=0}^{r=\infty}=a(0)=n \, ,
             \ee
             since $h(\infty)=1$. The topological charge of a $p$-Abelian Higgs model cannot take any other value but the vortex number $n$.}.
             
\subsubsection{Global charges}

We will also consider the energy $E$ and the angular momentum $J$ of these solutions. Their corresponding expressions can be derived from the $00$ and $0\vf$ components of the stress-energy tensor, respectively. Anticipating the model studied in Section {\bf 4}, the reduced one dimensional energy density of that model is
\bea
\label{ener_dens}
{\cal E}&=&\frac12\left(\frac{a'^2}{r^2}+b'^2\right)
+\tau\left(\frac{a^2}{r^2}+b^2\right) f'^2\sin^2 f 
\nonumber
\\
&&+\frac12\,\eta^2\,\left[f'^2 +\left(\frac{a^2}{r^2}+b^2\right)\sin^2f\right]+
\frac{1}{32}\,\la\,\left(\frac{\eta^3}{\ka}\right)^2\sin^2f\cos^2f\,.
\eea
The reduced angular momentum density is,
\be
\label{angmom_dens}
{\cal J} = a' b' + \eta^2 ab \sin^2 f + 2 \tau a b f'^2  \sin^2 f \,,
   \ee
   and the electric charge density is
   \be
\label{elechar_dens}
\rho_e = - \frac{1}{r} (r b')' - 4\kappa \frac{a'}{r}  \, ,
   \ee
   derived from the Gauss Law equation.
   
The definitions of both angular momentum and electric charge depend on the $0$-th component of the Abelian connection $A_0$, which in turn depends on the presence of the Chern-Simons term.

The energy, the angular momentum and the electric charge are then given by 
\be
\label{energy}
E = 2 \pi \int_0^\infty r {\cal E} dr  \,, 
\ee
\be
\label{angmom}
J = 2 \pi \int_0^\infty r {\cal J}  dr = 4\pi\kappa (a_\infty^2 - n^2)  \,. 
\ee
\be
\label{elechar}
Q_e =  2 \pi \int_0^\infty r \rho_e  dr = 8\pi\kappa (n - a_\infty)   \, . 
\ee
In contrast to our conclusion in Section {\bf 1.1.2} above, where we excluded the possibility of evolving topological charge of the Higgs model analogues, the situation here relating to \re{angmom} and \re{elechar} is different. This is because the definitions of ${\cal J}$ and of $\rho_e$ are independent of the scalar field content, $i.e.$, Skyrme, or, Higgs. Thus, as long as the Higgs model employed is a $p$-AH model with $p\ge 2$, which supports solutions with $a_\infty\neq 0$ like a Skyrme model, the evolutions of the spin $J$, \re{angmom}, and electric charge $Q_e$, \re{elechar}, can be tracked. In Ref.~\cite{Navarro-Lerida:2016omj} this was carried out in detail and the corresponding evolutions of the energy w.r. t. $Q$ and $J$ were tracked.

\section{$SO(2)$ gauged Skyrmions on $\R^2$ with Maxwell dynamics}
The Lagrangian and the static Hamiltonian of the $SO(2)$ gauged planar Skyrme model with Maxwell dynamics are
\bea
{\cal L}&=&-\frac14F_{\mu\nu}^2+\frac12\eta^2|D_\mu\f^a|^2-\eta^4\,V[\f^3]\ ,\quad \mu=0,i \, , \label{LMax}\\
{\cal H}&=&\frac14F_{ij}^2+\frac12\eta^2|D_i\f^a|^2+\eta^4\,V[\f^3]\ ,\quad i=1,2\,.\label{HMax}
\eea
Two potential functions $V[\f^3]$ in \re{LMax}-\re{HMax} will be considered.

Subjecting the static Hamiltonian \re{HMax} to the symmetry \re{axO3}-\re{Maxax},
the resulting one dimensional energy density functional is
\bea
\label{redHMax}
H&=&\frac{1}{2}\left[\frac{a'^2}{r}+\eta^2\left(r\,f'^2+\frac{a^2\sin^2f}{4r}\right)\right]+\eta^4\,V[\cos f] \, .
\eea
We study the vortices resulting from the use of the two potentials
\bea
V&=&\frac12\la(1-\f^3)^2\,,\label{pot1}\\
V&=&\la'(1-\f^3)\,.\label{pot2}
\eea

The potential \re{pot1}, with $\la=1$ leads to the saturation of the topological lower bound of the energy.
After imposition of symmetry on \re{bog1}-\re{bog2}, the two first order ODE's are
\bea
f'&=&-\frac{a\,\sin f}{r} \, , \label{Bog1}\\
\frac{a'}{r}&=&\eta^2\,(1-\cos f)\,.\label{Bog2}
\eea
This is the case studied in Ref.~\cite{Schroers:1995he}. Here we will consider vortices of models also with $\la\neq 1$, by solving the second-order equations of motion.
We will seek to find the energy profiles $versus$ the real and positive parameter $\la$. These profiles display discontinuities and branches.

The reason for considering the other potential, \re{pot2}, is that it is a rather natural choice resembling the ``pion mass'' potential
of the three dimensional Skyrme model. (The topological lower bound established \re{fHamMax} for the potential \re{pot1} remains valid in this case.)
Clearly with this choice of potential, only solutions to the second-order equations of motion can be sought. A distinguishing feature of these solutions is that
they decay exponentially, in contrast to the power decay in the case of potential \re{pot1}.

A common feature between the solutions of the models with potentials \re{pot1} and \re{pot2} is, that in both cases the topological charge equals the winding number $n$, as seen
from \re{topchv} with $v=1$.

We are specially interested here in solutions with the ranges of $\la$ which
are subject to the lower bounds \re{JR1}-\re{JR1}. Following Ref.~\cite{Jacobs:1978ch}, one can
estimate numerically the relative magnitude of the energy $per\ unit$ vorticity $n$.

\subsection{The solutions}
Assuming power behaviour in the $r\ll 1$ region, we find
\bea
f(r)&=&\pi+f_nr^{|n|}+o(r^{|n|+2}) \, ,\label{fat0}\\
a(r)&=&n+a_2r^2+o(r^4) \, , \label{bat0}
\eea
where $n$ is a non-zero integer.

At infinity, the value $f(\infty)=0$ is fixed by \re{bvf1}.
How $f(r)$ decays at infinity depends on the potential employed.

\medskip
{\bf a1:} The choice of potential $V=\frac12\la(1-\f^3)^2$, \re{pot1}, leads to a power decay, given by
\bea
f(r)&=&\sqrt{\frac{2}{\lambda}}\frac{1}{r} + \dots \, , \label{fatinfquad0}\\
a(r)&=& \frac{a_1}{r^{\sqrt{1+2/\lambda}-1}}+ \dots \, , \label{batinfquad0}
\eea
for $a_\infty = 0$, and
\bea
f(r)&=& \frac{f_1}{r^{a_\infty}}+ \dots \, , \label{fatinfquad}\\
a(r)&=& a_\infty + \frac{1}{4}\eta^2\frac{f_1^2}{a_\infty-1} \frac{1}{r^{2(a_\infty-1})}+ \dots \, , \label{batinfquad}
\eea
for $a_\infty \neq 0$.
For the quadratic potential \re{pot1} our numerical resolution is lower. We have obtained solutions for $\lambda<1$ which have $a_\infty=0$ and self-dual solution for $\lambda=1$ which have $a_\infty\neq0$ \cite{Schroers:1995he}.
The dependence of the energy on $\lambda$ for $\lambda \leq 1$ is shown in Figure \ref{fig_E_vs_lam_quadratic}. We see that in that region the interaction between vortices is attractive. For $\lambda>1$ the numerical schemes seem to indicate that the solutions might not exist, since their accuracy degenerates quite fast as higher resolution is required. Those solutions do exist if a Skyrme term is added to the theory, though. One can produce solutions with non-vanishing $a_\infty$ for any value of $\lambda$ when the Skyrme term is present, but as soon as one tries to remove it the convergence of the numerical schemes ceases, indicating that the solutions in the limit without the Skyrme term might not exist for $\lambda>1$.

\medskip
{\bf a2:} The choice of potential $V=\la'(1-\f^3)$, \re{pot2}, results in the exponential decay, given by
\bea
f(r)&=&\frac{f_1}{\sqrt{r}} e^{-\eta \sqrt{\lambda'}r} + \dots \, , \label{fatinflin}\\
a(r)&=& a_\infty + \frac{a_\infty f_1^2}{4\lambda'r}e^{-2\eta \sqrt{\lambda'}r} + \dots \, . \label{batinflin}
\eea

Both in \re{batinfquad} and \re{batinflin},
the value of $a_\infty$ on the other hand is in general fixed by the numerical process.

\begin{figure}[h!]
\centering
\includegraphics[height=3in]{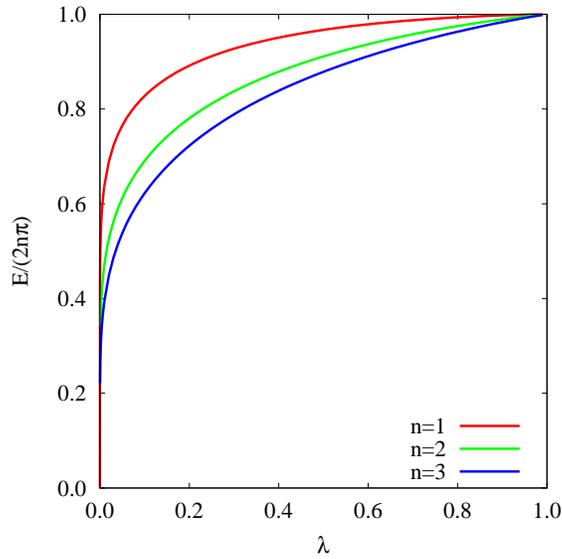}  
\caption{$E$ vs. $\lambda$ for (M) vortices with $n=1,2,3$, $\eta=1/\sqrt{2}$ for the 'quadratic' potential.} 
\label{fig_E_vs_lam_quadratic}
\end{figure} 

In Figure \ref{fig_E_a_inf_vs_lam_prime_linear} (left) we show the energy per vorticity $E/n$ for the vortices with the pion-mass potential \re{pot2}. We observe there is a limiting value of the coupling constant of the potential given by 
\be
\lambda'=\frac{n-1}{n} \, .
\ee
For $n=1$ the limiting value coincides with the minimal value of $\lambda'$ (zero) but for $n>1$ it corresponds to the maximal value of $\lambda'$. For these solutions $a_\infty$ ranges for 0 to $n$. The dependence of $a_\infty$ on $\lambda'$ is exhibited in Figure \ref{fig_E_a_inf_vs_lam_prime_linear} (right).

\begin{figure}[h!]
\centering
\includegraphics[height=3in]{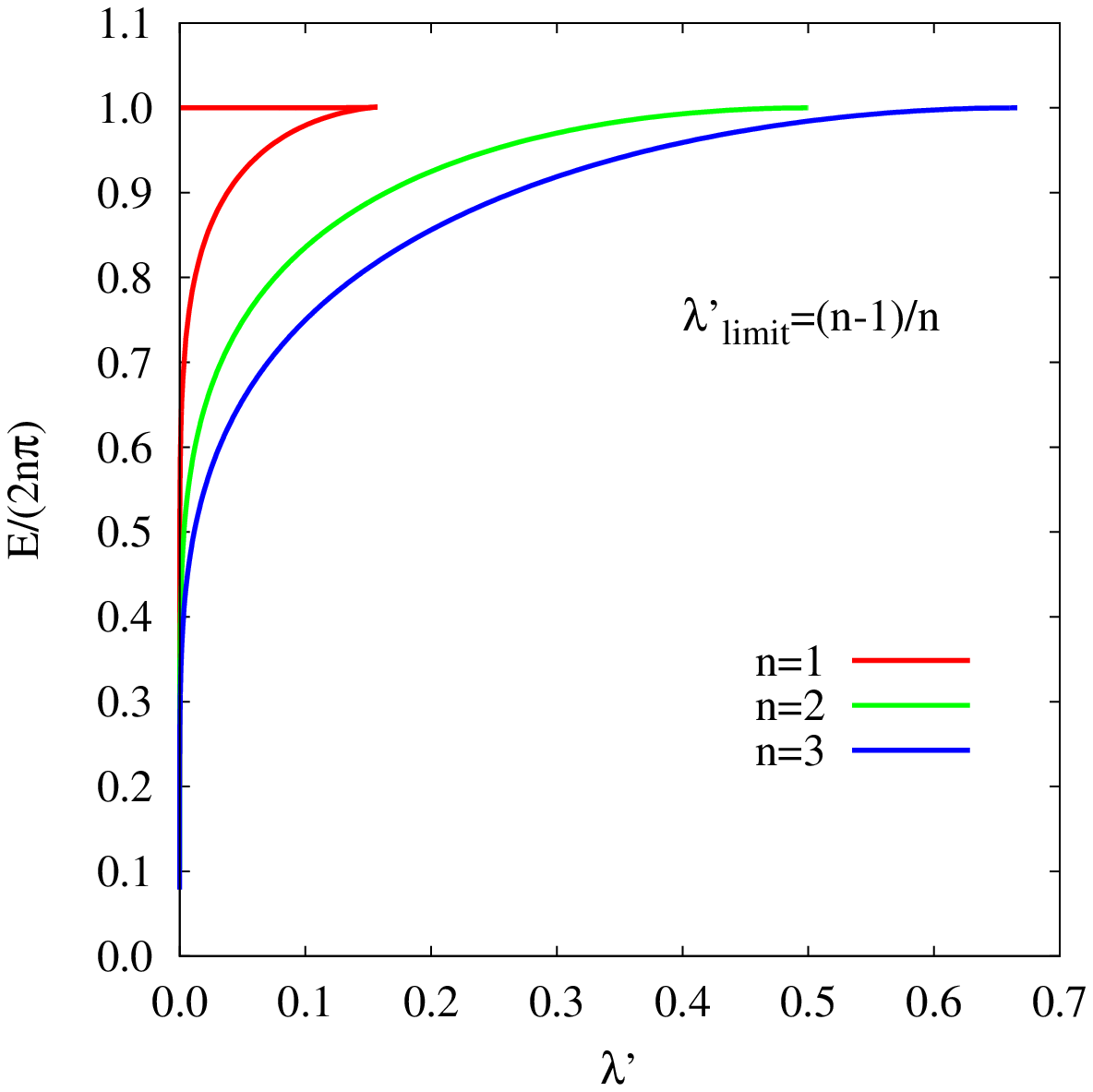}
\includegraphics[height=3in]{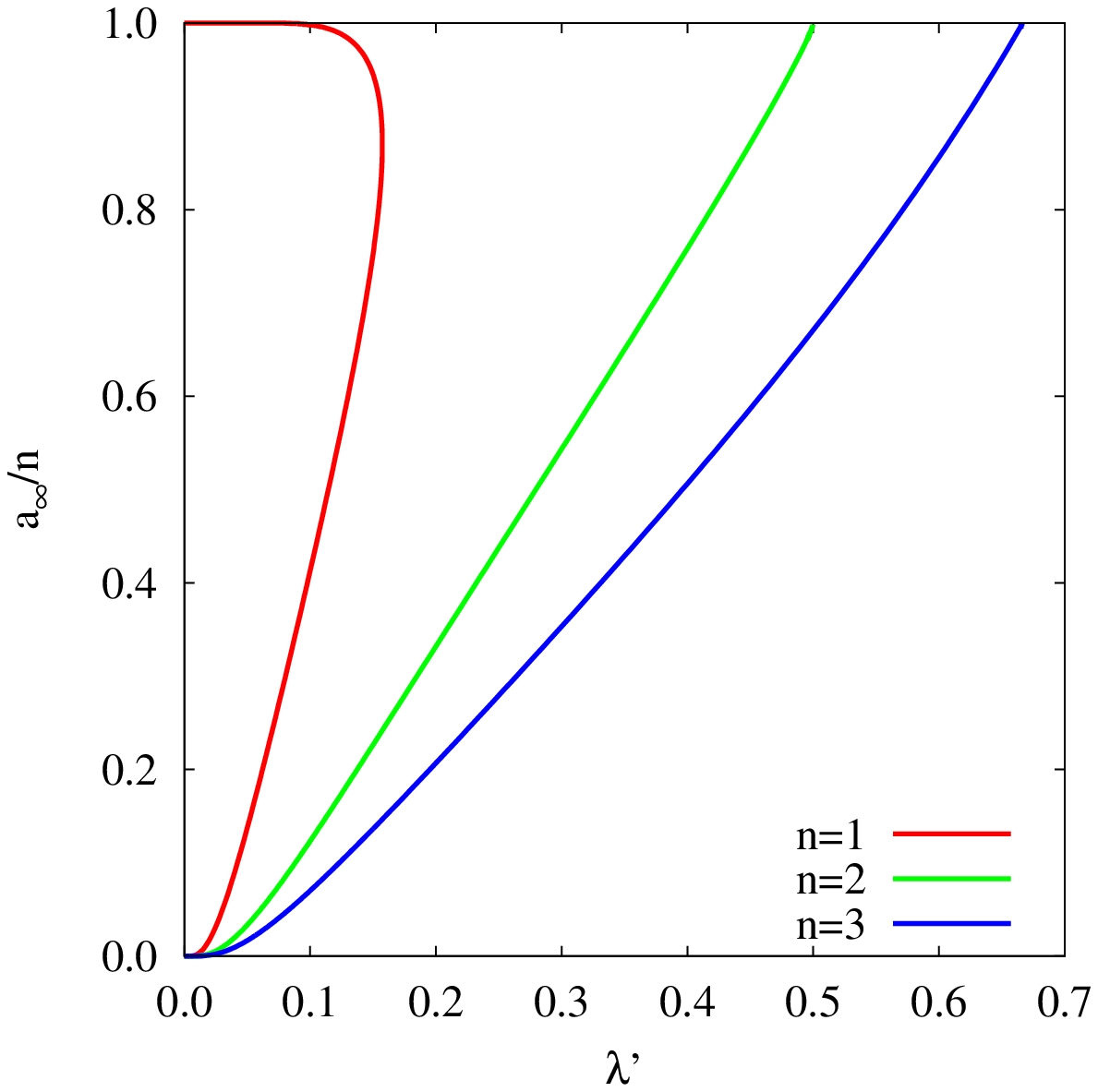}    
\caption{ {\it Left panel:}
$E$ vs. $\lambda'$ for (M) vortices with $n=1,2,3$, $\eta=1/\sqrt{2}$ for the 'pion-mass' potential.
{\it Right panel:}
$a_\infty$ vs. $\lambda'$ for the same vortices.} 
\label{fig_E_a_inf_vs_lam_prime_linear}
\end{figure} 



\section{$SO(2)$ gauged Skyrmions with Chern-Simons dynamics in $2+1$ dimensions}
Apart from its intrinsic interest, the main motivation for considering models whose Abelian field dynamics is
described by Chern-Simons term in the Lagrangian, is that it enables the choice of a potential in which the constant
$v$ may take values $v\neq 1$. This potential is given in Appendix {\bf C} by \re{VCS} and the topological lower bound
resulting from the Belavin inequality \re{Bfinineq} remains valid for any value of $v$. In the present work, we opt
for the simplest such value $v=0$. As a result, the topological charge defined by \re{9a} in Appendix {\bf A}, with 
$v=0$, results in the expression \re{fintopch} for the topological charge.

Equation \re{fintopch} enables the tracking the evolution of the topological charge, by tracking the evolving
values of $a_{\infty}$ that parametrise the solutions. We will find that for the model in this section the values of
$a_{\infty}$ do not result in the dissipation of the topological charge, so we defer this question to the study of the
model with both Chern-Simons and Maxwell dynamics in the next section. In the present section, we restrict our
considerations to the intrinsic properties of the solutions.

The Lagrangian and its static Hamiltonian studied in this section are defined by \re{CSLag} and \re{CSHamfin}, respectively,
in both of which the potential $V[\f]$ is given by \re{JRCS}. The static (energy) density employed in this model is
\re{CSHamfin}, which is seen from \re{Bfinineq} to be bounded from below by the topological charge. The static Hamiltonian is
\be
\label{bfinineq}
{\cal H}_{(\la)}=4\left(\frac{\ka}{\eta}\right)^2\frac{F_{ij}^2}{|\f^\al|^2}+\frac12\eta^2|D_i\f^a|^2+
\frac{\la}{32}\left(\frac{\eta^3}{\ka}\right)^2|\f^\al|^2(\f^3-\upsilon)^2\,.
\ee
What is different here, contrasting with the Maxwell case above, is that the value of the real constant $v$ is not restricted.
This is obvious since the potential term in \re{bfinineq} would vanish at infinity for any value of $v$
since $|\f^\al|^2(\infty)=0$. 

Another striking difference here, contrasting with the Maxwell case above, is
that in this case the second-order equations of motion are solved by the first-order equations \re{bogCS1}-\re{bogCS2}
for any value of $v$ for the self-dual case $\lambda=1$. As seen from \re{9a}, only when the choice $v=1$ is made, is the topological charge equal to the 
winding number $n$. For all other values of $v\neq 1$ the topological charge of the gauged system departs from $n$ and depends on $a_\infty$.

After imposition of symmetry using the Ansatz \re{axO3}-\re{Maxax}, \re{bfinineq} reduces to the one-dimensional density
\be
\label{redbfinineq}
H_{(\la)}=8\left(\frac{\ka}{\eta}\right)^2\frac{a'^2}{r\sin^2f}+\frac12\eta^2\left(r\,f'^2+\frac{a^2\sin^2f}{4r}\right)
+\frac{\la}{32}\left(\frac{\eta^3}{\ka}\right)^2r\sin^2f\,(\cos f-\upsilon)^2\,.
\ee
When $\la=1$ the second-order equations of motion are solved by the first-order equations
\bea
f'&=&\frac{a\sin f}{r} \,,\label{redsd1}\\
\ka\,a'&=&-2r\,U\,\sin^2f\,,\label{redsd2}
\eea
where the function $U$ in \re{redsd2}, given by \re{UCS}, reduces to
\be
\label{UCSred}
U=\frac18\left(\frac{\eta}{\ka}\right)(\cos f-\upsilon)\,.
\ee
These self-dual solutions were constructed in \cite{Arthur:1996uu}. Here, we seek solutions to the second-order equations, with $\la\neq 1$.

We will restrict our quantitative considerations to the models with $v=0$ since this is the simplest case of $v\neq 1$, this being the case of interest from the viewpoint of tracking the
evolution of the topological charge. The model with $v=1$ has been studied in Ref.~\cite{Arthur:1996nq}~\footnote{The emphasis there was in the
dependence of the energy per unit vortex number $n$ on the parameter $\la$. In comparison with the Abelian Higgs model,
studied in  Ref.~\cite{Jacobs:1978ch}, the results were qualitatively similar. Unlike the latter however, where the
solutions had boundary value $a_\infty=0$, in this model a continuous family of solutions parametrised by $a_\infty\neq 0$
was constructed.}.

Proceeding with the study of the $v=0$ model, the expansions at the origin for this solutions read
\bea
a(r) &=& n + \frac{\eta^2 b_0 f_n^2}{8(n+1)\kappa} r^{2(n+1)} + \dots \, , \label{CS_Sk_a_at_or} \\
b(r) &=& b_0 + \frac{\eta^2f_n^2}{8\kappa}r^{2n} + \dots \, , \label{CS_Sk_b_at_or} \\
f(r) &=& \pi + f_n r^n + \dots \, . \label{CS_Sk_f_at_or}
\eea

Concerning the asymptotic behaviour, the situation is more complicated. We have to distinguish between two cases: $a_\infty>n$ and $a_\infty < n$ ($a_\infty = n$ is a singular case). For $a_\infty>n$ we have a power decay in a $1/r$ resulting to be
\bea
a(r) &=& a_\infty - \frac{1}{128}\frac{\lambda}{a_\infty-1}\frac{\eta^6}{\kappa^3} f_1^2 \frac{1}{r^{2(a_\infty-1)}} + \dots  \, ,\label{CS_Sk_a_at_inf} \\
b(r) &=&\frac{1}{16}\frac{\eta^3}{\kappa^2}\lambda - \frac{1}{8\kappa}\eta^2f_1^2\frac{1}{r^{2a_\infty}} + \dots \, ,\label{CS_Sk_b_at_inf} \\
f(r) &=&\frac{f_1}{r^{a_\infty}}  + \dots \, . \label{CS_Sk_f_at_inf}
\eea

For $a_\infty < n$ the functions decay exponentially. The corresponding expressions are quite involved so we will not reproduce them here.

This change in the asymptotic behaviour is reflected on the dependence of the energy on $a_\infty$. In Figure \ref{fig_E_b_inf_vs_ainf_CS_Sk} (left) we represent the energy $E$ versus the parameter $a_\infty$ for $SO(2)$ gauged Skyrmions with CS coupling constant $\kappa=1$, $\lambda=1$ and vorticities $n=1,2,3,4$. For $a_\infty>n$ the energy becomes $E=2\pi(n+a_\infty)$, namely, it saturates the topological lower bound. Whereas for $a_\infty < n$ the dependence of the energy on $a_\infty$ is different. Such solutions, which may be considered to be sphalerons, were not sought in \cite{Arthur:1996nq} in the case of the model with $v=1$.

\begin{figure}[h!]
\centering
\includegraphics[height=3in]{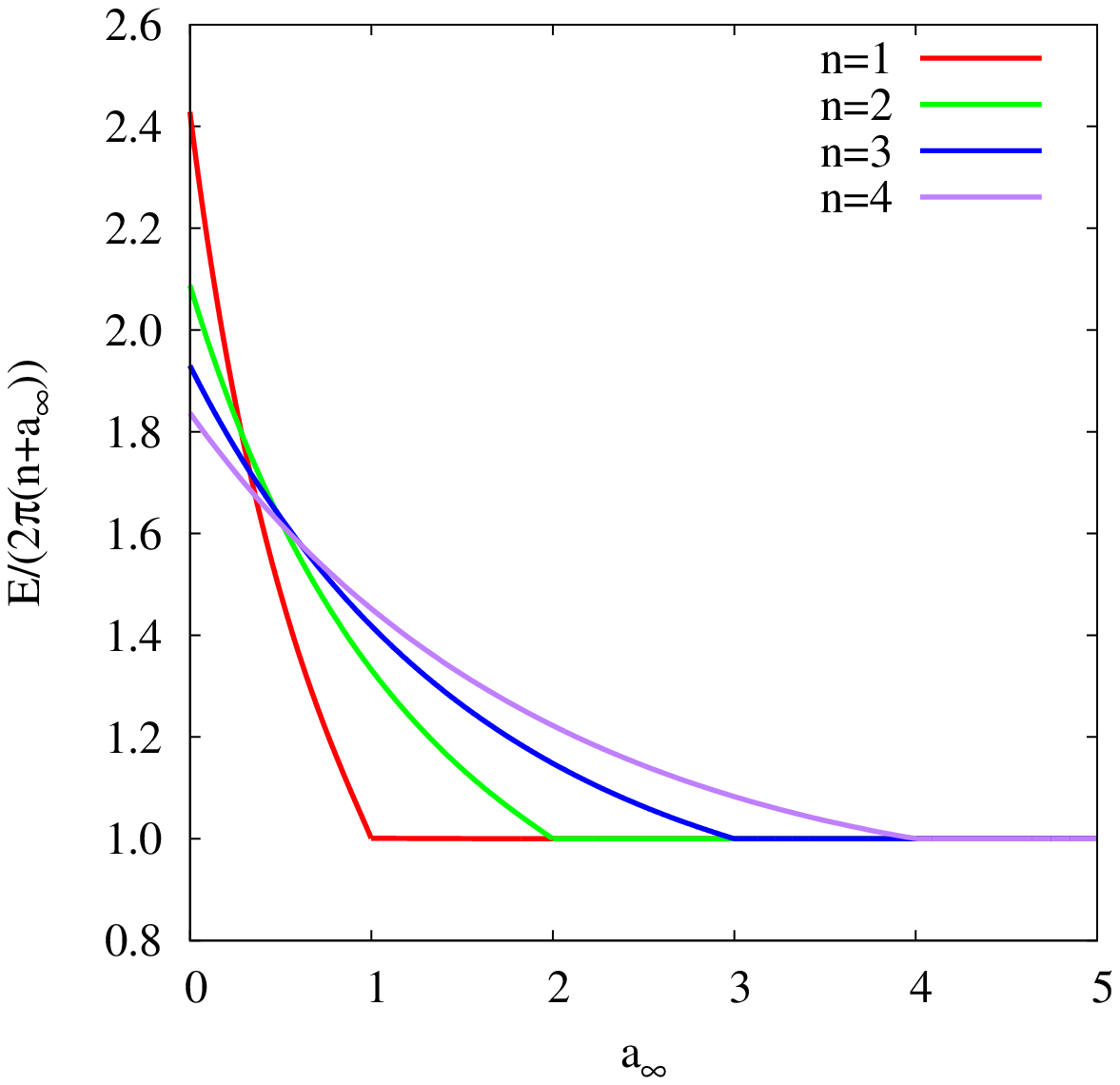}
\includegraphics[height=3in]{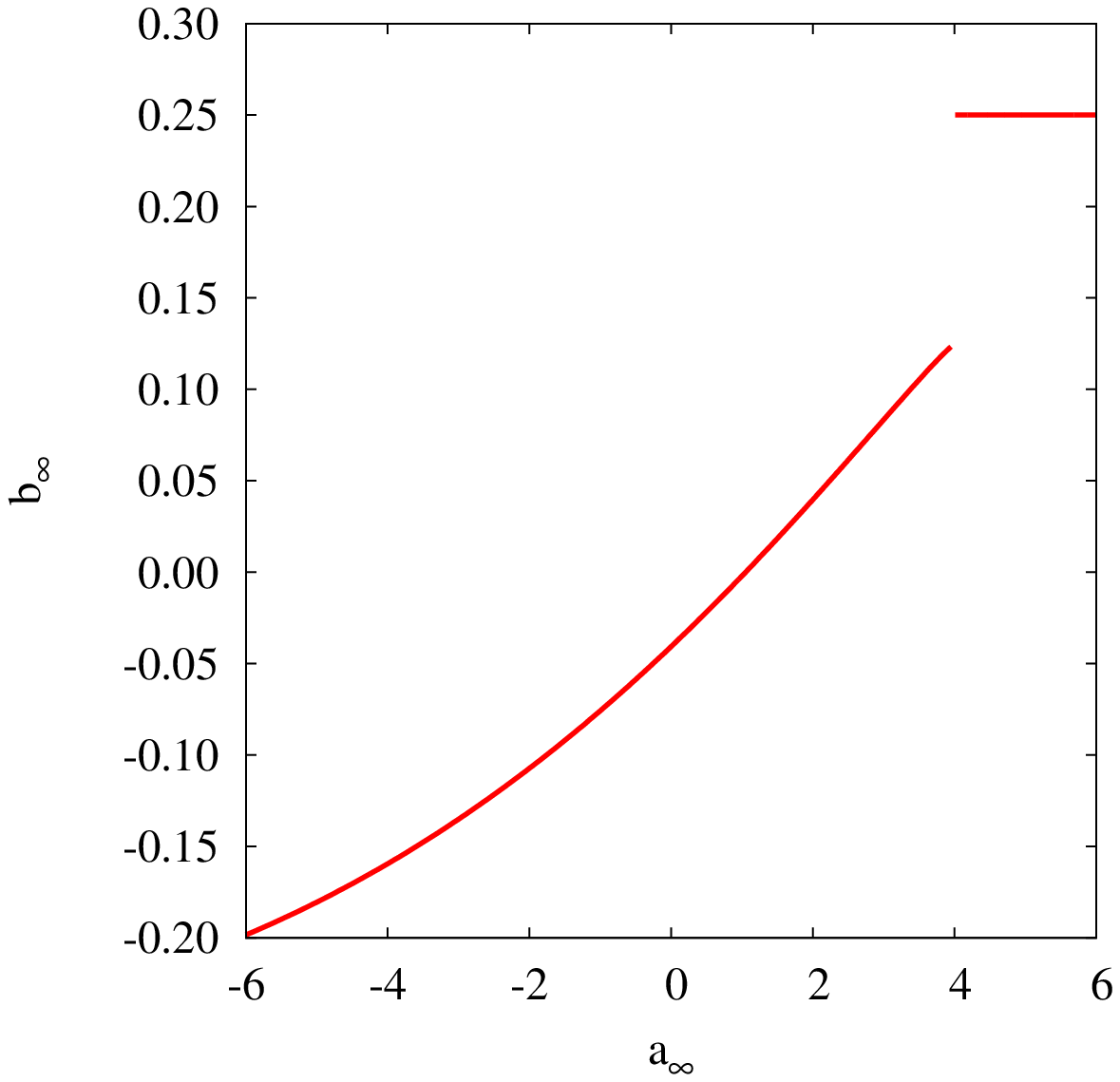}    
\caption{ {\it Left panel:}
$E$ vs. $a_\infty$ for (CS) vortices with $n=1,2,3,4$, $\eta=1$, $\kappa=1$ and $\lambda=1$.
{\it Right panel:}
$b_\infty$ vs. $a_\infty$ for (CS) vortices with $n=4$, $\eta=1$, $\kappa=1$ and $\lambda=1$.} 
\label{fig_E_b_inf_vs_ainf_CS_Sk}
\end{figure} 



Although the transitions from $a_\infty < n$ to $a_\infty > n$ look continuous at the level of the energy (with change in the slope at $a_\infty = n$), it is not so in fact. If one looks at the quantity $b_\infty$ we can observe a jump at $a_\infty=n$. We show this in Figure \ref{fig_E_b_inf_vs_ainf_CS_Sk} (right) for the set of solutions with $n=4$ shown in Figure \ref{fig_E_b_inf_vs_ainf_CS_Sk} (left). Notice that above $a_\infty=n$, $b_\infty$ is constant (and equal to the leading term of equation (\ref{CS_Sk_b_at_inf})).

We finally show the dependence of the energy on $\lambda$. That behaviour depends on the value of $a_\infty$ chosen. For $a_\infty<1$ we have the standard pattern where the curves for several vorticities do not cross each other, resulting in an attractive phase (see Figure \ref{fig_E_vs_lam_CS_Sk} for $a_\infty=0$). However, for $a_\infty>1$ solutions with $n<a_\infty$ end up crossing other curves.  We see that the situation here, for the pure Chern-Simons model with $v=0$ is much less transparent than in the corresponding model with $v=1$~\cite{Arthur:1996nq}, and in the corresponding Higgs model~\cite{Jacobs:1978ch}. We do not pursue this line further.

\begin{figure}[h!]
\centering
\includegraphics[height=3in]{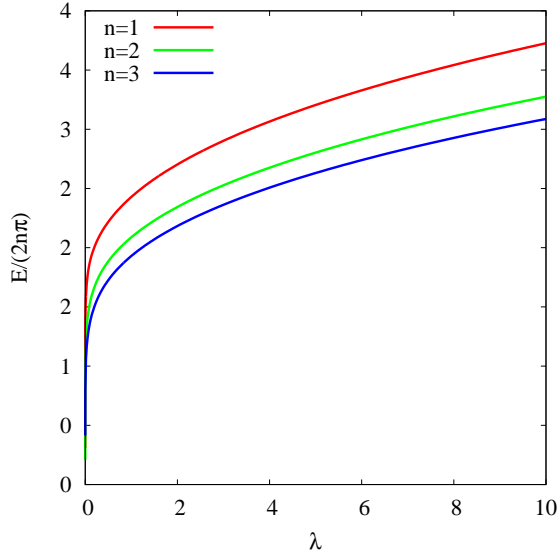}  
\caption{$E$ vs. $\lambda$ for (CS) vortices with $n=1,2,3$, $\eta=1$, $\kappa=1$ and $a_\infty=0$.} 
\label{fig_E_vs_lam_CS_Sk}
\end{figure} 


\section{$SO(2)$ gauged Skyrmions with Maxwell--Chern-Simons
dynamics in $2+1$ dimensions}
While in the previous two sections our attention was focused on the energy profiles versus $\la$,
the dimensionless constant parametrising the coupling
of the potential, here instead we focus on the evolution of the topological
charge in a given Maxwell--Chern-Simons--Skyrme theory. The topological lower bound for this model follows
immediately from the topological lower bound \re{Bfinineq} given in Appendix {\bf C} for the model with Chern-Simons
dynamics only, since adding the Maxwell term to the Lagrangian does not invalidate the ensuing
topological lower bound \re{Bfinineq}. In particular as elsewhere in this work, we will select the simplest option $v\neq1$, namely $v=0$.

As can be seen from the Lagrangian \re{CSLag}, it does not contain a quartic (``Skyrme'') kinetic term. We add such
a term nonetheless, so that in the gauge decoupling limit a topological lower bound persists. The soliton of the
ungauged system is stabilised by the ``baryon number'', the departure of the topological charge from which, is
what we aim to track.

We will find out in fact, that the evolution of the solutions can result in the total annihilation of the topological
charge.

The model is described by the Lagrangian
\be
\label{MCSLag}
{\cal L}_{(\la)}=-\frac14F_{\mu\nu}^2+\ka\vep^{\la\mu\nu}A_\la F_{\mu\nu}
-\frac18\tau D_{[\mu}\f^aD_{\nu]}\f^b+\frac12\eta^2|D_\mu\f^a|^2-\eta^4\,V[\f] \, ,
\ee
in which the potential $V[\f]$ is that given by \re{VCS}, with $v=0$.

This choice of $V[\f]$ is predicated by our desire to avail of the topological lower bound \re{Bfinineq},
where the charge density is that given by
\re{9a}, or, by \re{ttp}. This topological lower bound is not invalidated by adding the
Maxwell term to the Lagrangian \re{CSLag} which results in the addition of
a positive definite contribution to the static density \re{CSHam}.

While in the absence of the Maxwell term the electric component of the Abelian potential $A_0$ could be solved
using the Gauss Law equation, \re{Gauss} in Appendix {\bf C}, here this is not possible. So in the imposition of symmetry,
the function $b(r)$ in \re{Maxax0} appears explicitly in the one-dimensional reduced Lagrangian, 
which results to be
\bea
\label{redMCSLag}
-r^{-1}\,L&=&\frac12\left(\frac{a'^2}{r^2}-b'^2\right)
-\frac{2\ka}{r}[(ab'-ba')-nb']
+\tau\left(\frac{a^2}{r^2}-b^2\right) f'^2\sin^2 f 
\nonumber
\\
&&+\frac12\,\eta^2\,\left[f'^2 +\left(\frac{a^2}{r^2}-b^2\right)\sin^2f\right]+
\frac{1}{32}\,\la\,\left(\frac{\eta^3}{\ka}\right)^2\sin^2f\cos^2f\,,
\eea
in which the potential term is multiplied by the real positive parameter $\la$, in order to allow for generic values of the potential coupling constant.


The second-order equations of \re{redMCSLag} are solve subject to the boundary values
\bea
&&\lim_{r\to 0}f(r)=\pi\ ,\ \quad\lim_{r\to 0}a(r)=n\ ,\ \quad\lim_{r\to 0}b'(r)=0 \ ,\label{sk0}\\
&&\lim_{r\to\infty}f(r)=0\ ,\quad\lim_{r\to\infty}a(r)=a_\infty,\ ,\quad\lim_{r\to\infty}b(r)=b_\infty \ , \label{skinfty}
\eea
where $a_\infty$ is not necessarily zero and $b_\infty$ is a free parameter that allows us to vary the electric charge of the solutions within a concrete model (i.e., choice of the parameters in the Lagrangian).
Notice that $a_\infty$ is numerically related to $b_\infty$.

\subsection{Numerical results}

The field equations resulting from (\ref{redMCSLag}) cannot be integrated analytically and one has to resort to numerical methods to carry out that task. Given a concrete theory (e.g. fixed $\kappa$, $\tau$, $\eta$, and $\lambda$) and a fixed winding number $n$, there is just one free parameter to be varied. As said above we use $b_\infty$ as such a parameter, $a_\infty$ being numerically related to $b_\infty$. In Fig.~\ref{fig_binf_vs_ainf} we show the relation between $b_\infty$ and $a_\infty$ for a generic solution ($n=1$, $\lambda=1.6$, $\eta=1$, $\kappa=1$, and $\tau=1$). We observe that $b_\infty$ is bounded while $a_\infty$ seems not to be \footnote{We have found certain regions in the parameter space where $a_\infty$ could not be increased or decreased arbitrarily. However, it was not clear whether that fact was due to the actual absence of solutions or to a difficulty of the methods used to find them.}. That indicates that there exist solutions with arbitrarily large (positive or negative) values of the topological charge $q$. The same holds for the electric charge $Q_e$ and the angular momentum $J$. Concerning the energy $E$, it diverges as $|a_\infty|$ gets larger and larger. In fact, we know already from our results in Ref.~\cite{Navarro-Lerida:2016omj} that due to the dependence between $a_\infty$ and $b_\infty$, the dependence of the mass/energy $E$ on the global charges is non-standard, namely the dependence on the electric charge
$Q_e$ and the angular momentum $J$ displayed respectively (left and right panels) in Figure~\ref{figs2}.

\begin{figure}[h!]
\centering
\includegraphics[height=3in]{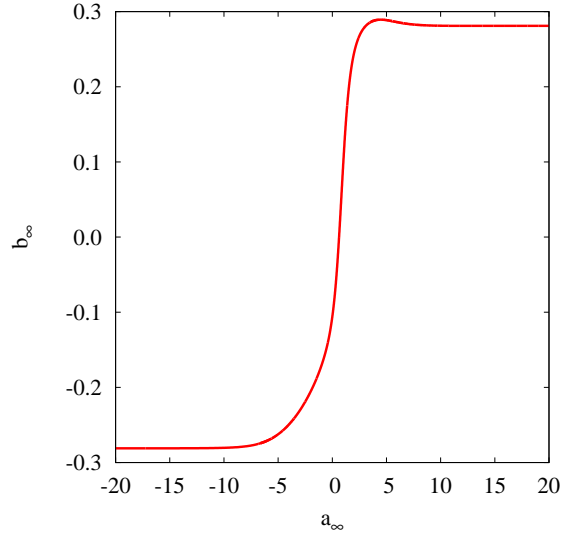}  
\caption{$b_\infty$ vs. $a_\infty$ for (MCS) vortices with $n=1$, $\lambda=1.6$, $\eta=1$, $\kappa=1$, and $\tau=1$.} 
\label{fig_binf_vs_ainf}
\end{figure} 

\begin{figure}[h!]
\begin{center}
\includegraphics[height=3in]{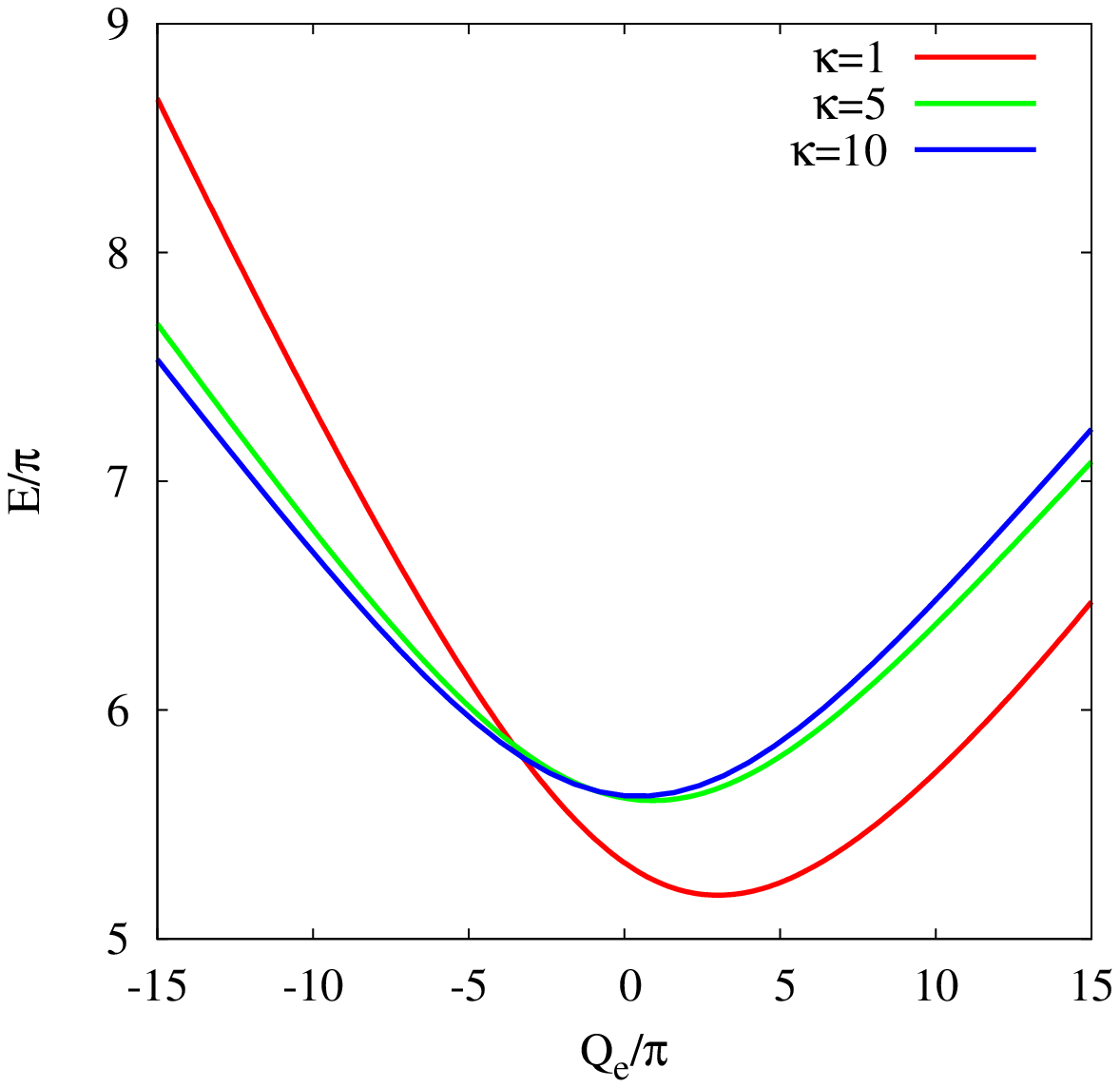}
\includegraphics[height=3in]{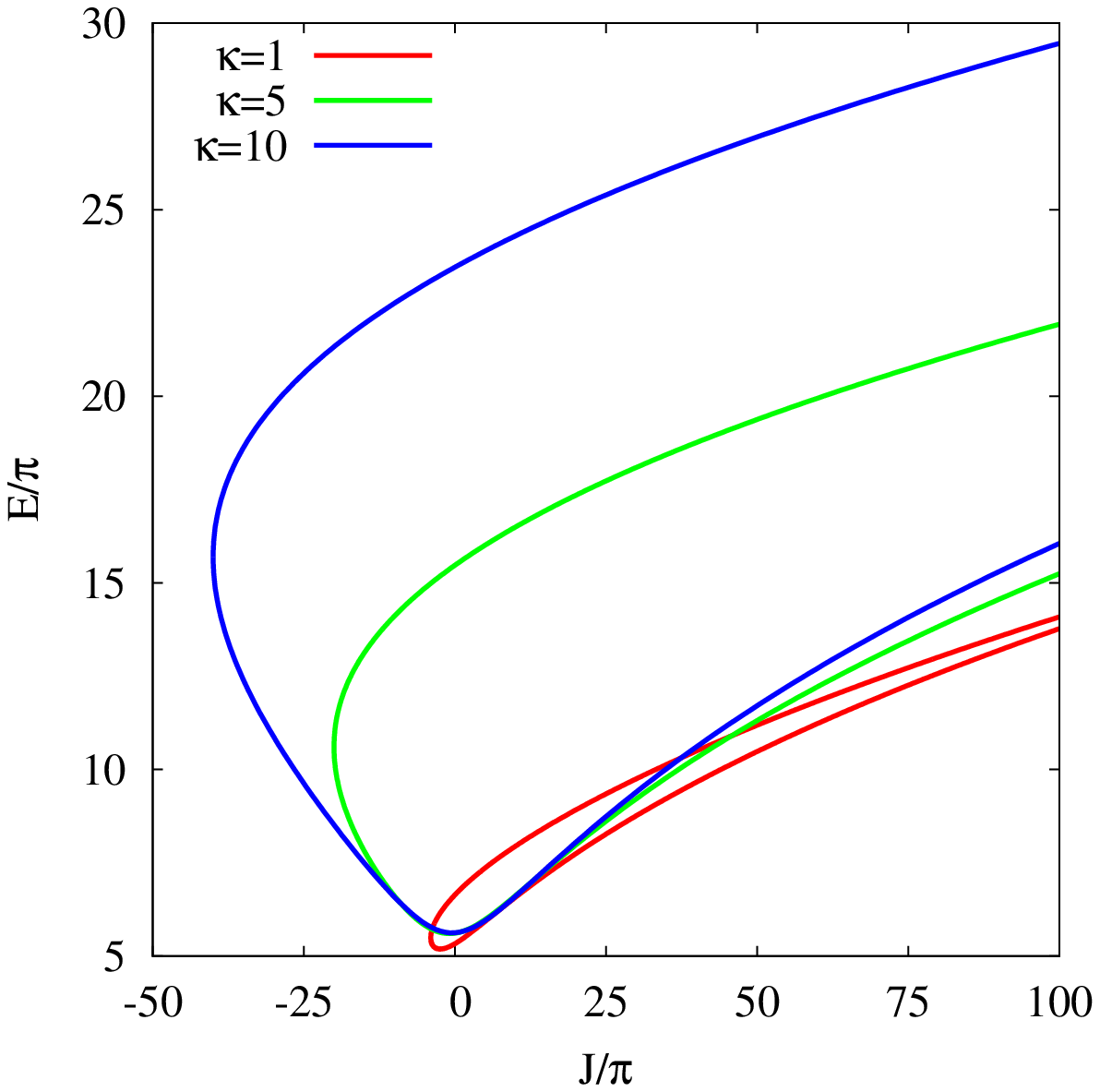}    
\caption{ {\it Left panel:} Energy $E$ $vs.$ electric charge $Q_e$ for (MCS) vortices with $n=1$, $\lambda=1.6$, $\eta=1$, $\tau=1$, and $\ka=1, 5, 10$. {\it Right panel:} Energy $E$ $vs.$ angular momentum $J$  for the same vortices.} 
\label{figs2}
\end{center}
\end{figure} 

\begin{figure}[h!]
\centering
\includegraphics[height=3in]{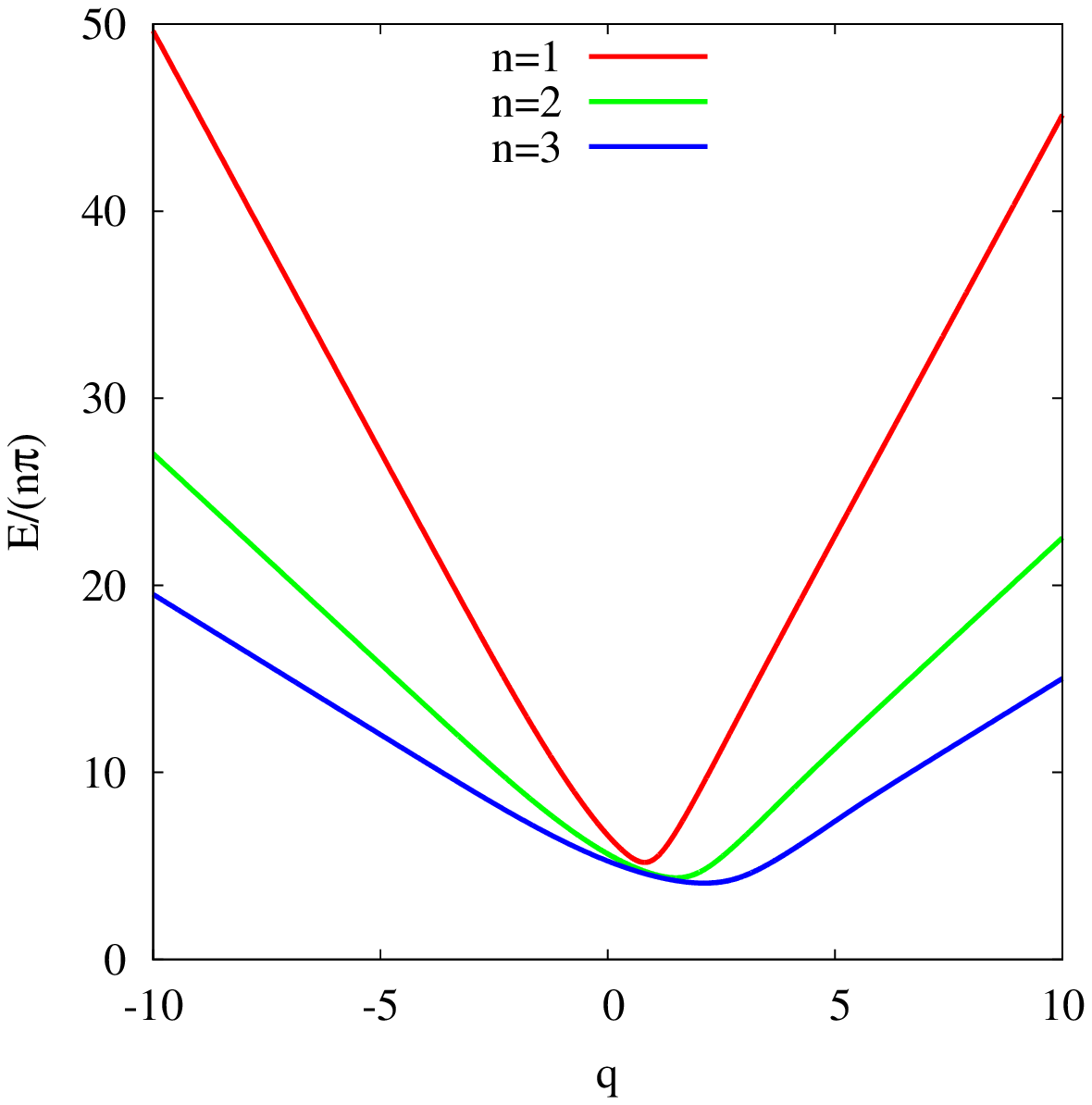}
\includegraphics[height=3in]{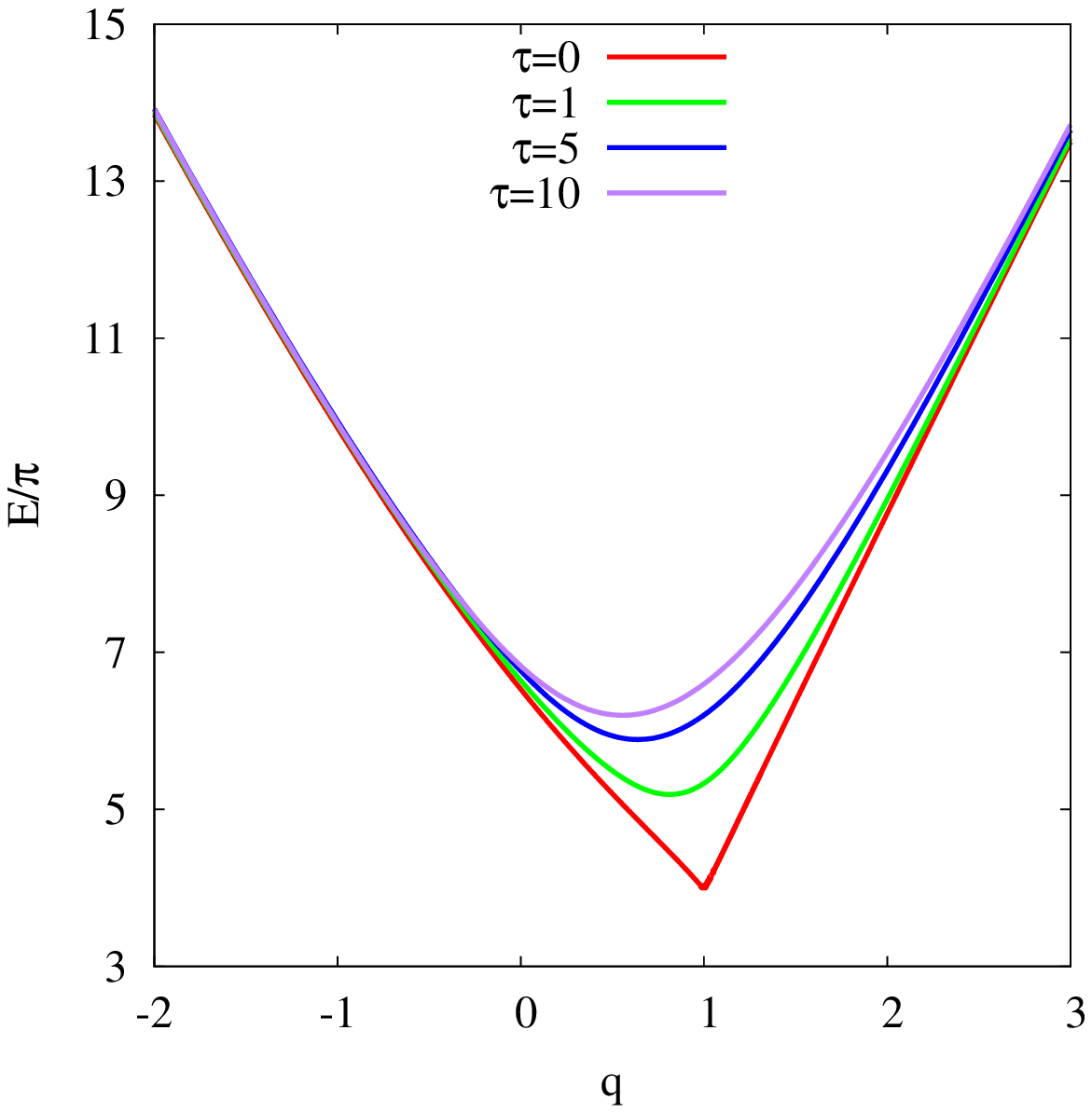}    
\caption{ {\it Left panel:}
$E$ vs. $q$ for (MCS) vortices with $n=1,2,3$, $\lambda=1.6$, $\eta=1$, $\kappa=1$, and $\tau=1$.
{\it Right panel:}
Same for (MCS) vortices with $n=1$, $\lambda=1.6$, $\eta=1$, $\kappa=1$, and $\tau=0,1,5,10$.} 
\label{fig_E_vs_q_sev_n}
\end{figure} 



Let us analyze the effect of the topological charge $q$ on the energy $E$. In Figure \ref{fig_E_vs_q_sev_n} (left) we represent the energy per unit winding number $n$ versus the topological charge for three values of $n$ for the same parameters as in Figure \ref{fig_binf_vs_ainf}. For small values of the Skyrme constant $\tau$ the minimum of the energy occurs at values of the topological charge around the winding number. However, when higher values of $\tau$ are considered the minimum of the energy occurs at values clearly different from the winding number. This is shown in Figure \ref{fig_E_vs_q_sev_n} (right), where the minimum of the curve for $n=1$, $\tau=10$ is located at $q \approx 0.556$. In the limit of large $\tau$ the minimum of the energy occurs at $q=n/2$, as shown in Figure \ref{fig_E_vs_q_sev_n_tau_1000}. So the minimum takes place within the range $q\in[n/2,n]$.  Concerning the stability of these solutions, one would be tempted to state that most stable solution would correspond to that with less energy, which according to Figure \ref{fig_E_vs_q_sev_n} (right) does not possess integer topological charge, in general.

\begin{figure}[h!]
\centering
\includegraphics[height=3in]{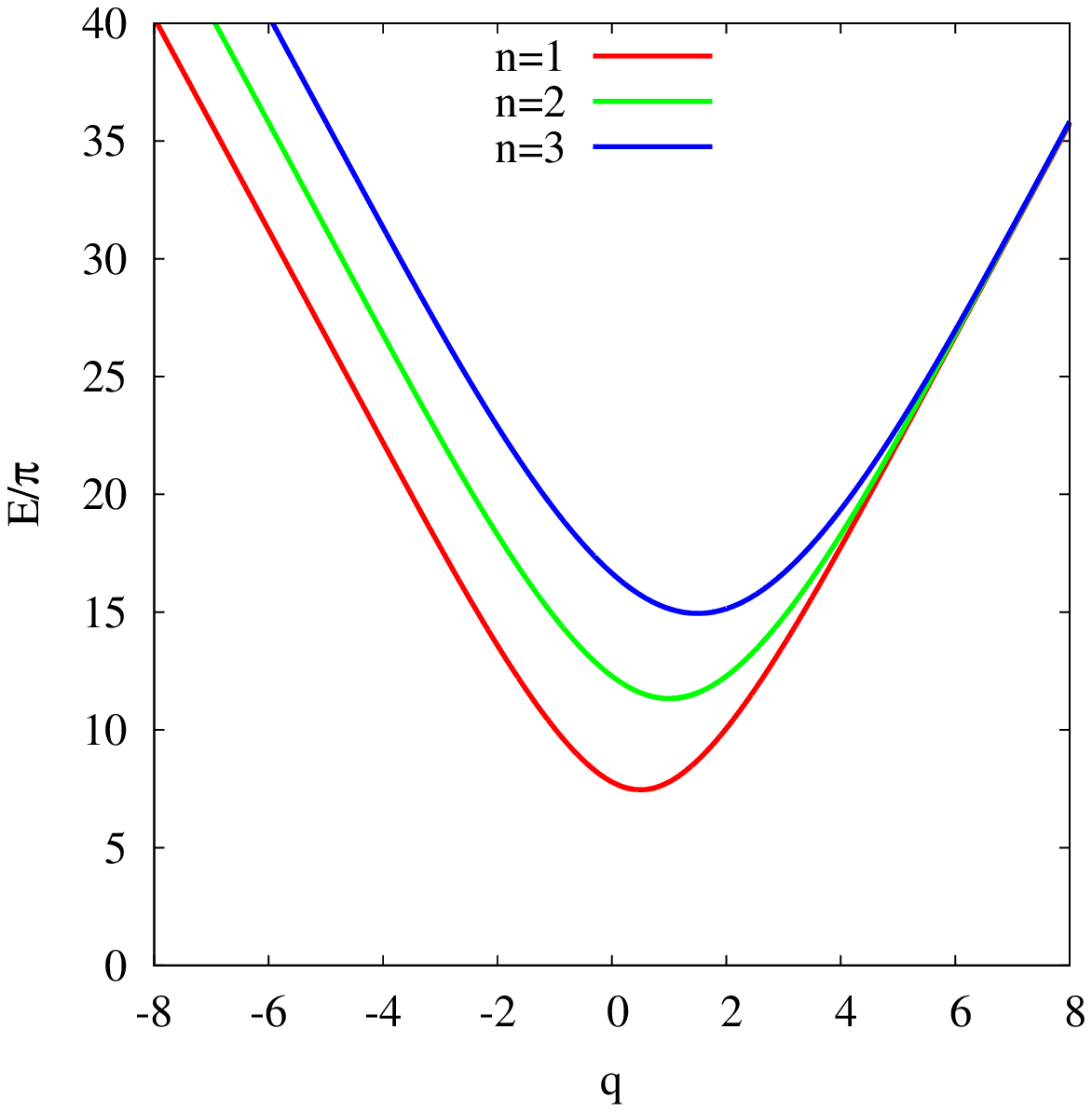}  
\caption{$E$ vs. $q$ for (MCS) vortices with $n=1,2,3$, $\lambda=1.6$, $\eta=1$, $\kappa=1$, and $\tau=1000$.} 
\label{fig_E_vs_q_sev_n_tau_1000}
\end{figure} 



\section{Summary and discussion}
We have made a systematic study of $SO(2)$ gauged planar Skyrmions in $2+1$ dimensions. The models in question feature {\bf a)} Maxwell (only), {\bf b)} Chern-Simons (only) and {\bf c)} (both) Maxwell and Chern-Simons dynamics.
The studies in {\bf a)} and {\bf b)} are aimed at exposing $discontinuities$ and $branchings$ in the energy profiles of such solutions, which were encountered in the solutions to the $SO(3)$ gauged $O(4)$ Skyrme model on $\R^3$.
These were shown in Refs.~\cite{Brihaye:1998vr,Brihaye} and \cite{Kleihaus:1999ea},
and are expected to be typical features of all gauged sigma model, in contrast with (gauged) Higgs models where energy profiles are always $continuous$ and exhibit $monotonic$ behaviour.

In case {\bf a)}, pertaining to Maxwell dynamics (only), we have studied the model {\bf a1)} with  potential \re{pot1} that supports self-dual solutions of first-order equations, and the model {\bf a2)} with a ``pion mass'' potential
which has solutions to the second-order equations only. The energy profiles of both models exhibit non-monotonic features.

The solutions we found in the model {\bf a1)} are characterised by $a_\infty=0$, for the coupling constant $\la<1$, $\la=1$ corresponding to the self-dual solution. However, when
the model is augmented with the $quartic$ (Skyrme) kinetic term ($\tau>\neq 0$), we find solutions with any $\la$ and $a_\infty\neq 0$. Not surprisingly the quartic term, which has a smoothing effect, becomes singular in the limit $\tau\to 0$.
The energy profiles for $\tau=0$ are given in Figure \ref{fig_E_vs_lam_quadratic}, which while they do not exhibit branches and look smooth, they do stop at $\la=1$.

The solutions of the model {\bf a2)} with potential \re{pot2} exhibit more marked discontinuities in the energy profiles, which range from zero to some finite value of the coupling $\la'$ as shown in Figure \ref{fig_E_a_inf_vs_lam_prime_linear} (left) for $n=1,2,3$.
The behaviour of these profiles is typified by the value of $a_\infty$, which emerges from the numerical process. These latter are exhibited in Figure \ref{fig_E_a_inf_vs_lam_prime_linear} (right). The end point of each of these profiles is observed to coincide with the value of the coupling
$$\la'=\frac{n-1}{n}.$$

In case {\bf b)}, pertaining to Chern-Simons dynamics (only), we have studied only one model. This model, typified by a potential term exhibiting no constant $i.e.$ $v=0$ in \re{bfinineq}. This choice was made in anticipation of selecting the desired model {\bf c)},
subsequently studied in Section {\bf 3}. The solutions here present a more complex pattern, occasioned by two different asymptotic behaviors characterised by $a_\infty>n$ and $a_\infty<n$. When $a_\infty>n$, the functions decay with a power behaviour, while when
$a_\infty<n$ they decay exponentially, solutions with $a_\infty=n$ being singular. It is noteworthy to point out that the exponentially decaying solutions with $a_\infty>n$, the energy has the constant value $$E=2\pi(n+a_\infty),$$ as seen from Figure \ref{fig_E_b_inf_vs_ainf_CS_Sk} (left).
While the transition from the $a_\infty<n$ to the $a_\infty>n$ solution looks smooth in this plot, it is in fact discontinuous, as seen from Figure \ref{fig_E_b_inf_vs_ainf_CS_Sk} (right) showing the relation of $a_\infty$ and $b_\infty$ for $n=4$, where the jump at $n=4$ is seen.

Concerning the energy profiles versus the coupling $\la$ in this case, {\bf b)}, we have plotted these in Figure \ref{fig_E_vs_lam_CS_Sk} for the solutions $a_\infty<n$. These $E/n$ profiles are smooth and they do not cross at the self-duality point $\la=1$, hence
the system describes solutions exclusively in the attractive phase. In this respect,
these solutions differ from those of the vortices of the Abelian Higgs model, where both attractive and repulsive phases occur, and also , from vortices~\cite{Arthur:1996nq} of the Chern-Simons--Skyrme model with $v=1$ rather than the $v=0$ model studied here.
   The situation is different for the $a_\infty>n$ solutions, in which case curves of different vorticity do seem to cross, or at least meet. This completes our $first\ task$, namely the
   investigation of discontinuities and branchings in the energy profiles of gauged (planar) Skyrmions,
employing models featuring only Maxwell and only Chern-Simons dynamics, in turn, and, in the absence of the quartic Skyrme kinetic term.

In carrying out our $second\ task$, namely the tracking of the topological charge with changing energy, we proceeded to
the third case {\bf c)}. The system studied contains both Maxwell and Chern-Simons
dynamics~\footnote{Such systems, {\bf c)}, studied here and in \cite{Navarro-Lerida:2016omj}, do not result in energy profiles exhibiting discontinuities and singularities. The solutions are even smoother, when the $quartic$ kinetic Skyrme term is added.}.
Such a system was studied in (Section {\bf 4} of) Ref.~\cite{Navarro-Lerida:2016omj}, albeit with a different (Skyrme) potential term.
There, the emphasis was on the non-standard dependence of the global charges, electric charge and spin, while here, the emphasis is on the evolution of the topological charge and its dependence on the energy. The choice of potential in the present work, is motivated
by our desire to influence the topological charge. As can be seen from the definition \re{9a} of topological charge, if $v=1$ as $e.g.$ in the ``pion mass'' potential, the integral of the charge density will yield the winding number $n$, as in the gauge-decoupled case.
Here, we have adopted a potential featuring the value $v=0$, which does not force the topological charge to take the value $n$. This
potential is inspired by Chern-Simons dynamics, arrived at via the Belavin inequalities presented in Appendix {\bf C}.

In the case {\bf c)} here, we have included also the $quartic$ (Skyrme) kinetic term with coupling strength $\tau$, since we require that our model support solitons~\cite{Battye:2005nx} in the gauge-decoupling limit.

The mechanism giving rise to the features found in \cite{Navarro-Lerida:2016omj} and here, is the same one, namely the effect of the Chern-Simons dynamics.
Technically, this is due to the intertwining of the magnetic and electric functions $a$ and $b$ present in the Chern-Simons density, the asymptotic values $a_\infty$ and $b_\infty$ of which characterise the solutions in any given model.
In the present case, the relation between $a_\infty$ and $b_\infty$ is displayed in Figure \ref{fig_binf_vs_ainf}, which corresponds to Figure {\bf 6} in Ref.~\cite{Navarro-Lerida:2016omj}. The latter result in the non-standard dependence of the energy on the global
charges featured in Figures {\bf 7} and in {\bf 8} in \cite{Navarro-Lerida:2016omj}. The model studied in \cite{Navarro-Lerida:2016omj} featured the ``pion-mass'' potential, while the potential used here is \re{VCS} (with $v=0$) employed in \re{MCSLag}.
For completeness, we reproduce these results for the system studied here, in Figures \ref{figs2} (left) and \ref{figs2} (right).

The most interesting result in the present work is the dependence of the topological charge on the energy. In Figures  \ref{fig_E_vs_q_sev_n} (left), \ref{fig_E_vs_q_sev_n} (right), and \ref{fig_E_vs_q_sev_n_tau_1000} we observe that the topological charge evolves, taking values different from the ``baryon number'' $n$, or the winding number, due to
the effect of the gauge field. We observe that in the absence of the quartic kinetic (Skyrme) term ($\tau=0$) the energetically favoured states occur at the value of the topological charge coinciding with the winding number. When the Skyrme term is switched on,
the energetically favoured states appear at smaller values, down to $one \ half$ of the winding number for large $\tau$. Also in the limit of large $\tau$, we find a clear asymmetry between positive and negative values of the topological charge $q$. For large positive values of $q$ the energy becomes independent of the winding number $n$, whereas for large negative values of $q$ the gap in energy between curves with consecutive values of $n$ does not depend en $n$, i.e., the curves are equispaced.

The present work is entirely devoted to the study of gauged Skyrmions in $2+1$ dimensions. Of special interest is the influence of Chern-Simons dynamics which results in the non-standard behaviour of the energy on global charges, as reported in
\cite{Navarro-Lerida:2016omj}. Going further, in the present work we investigate
the effect of the Chern-Simons dynamics on the evolution of the topological charge. We expect that these effects are not limited to the $2+1$ dimensional Skyrmions, but that they feature also in
higher dimensions.

In a recent Letter~\cite{Navarro-Lerida:2018giv}, $SO(2)$ gauged Skyrmions in both $2+1$ and $3+1$ dimensions were studied. It was found there that in the $3+1$ dimensional case, where there is no
usual~\footnote{Skyrme--Chern-Simons densities, proposed in \cite{Tchrakian:2015pka},
can be defined in even dimensions, but a systematic (numerical) study of this has not been attempted to date.} definition of a Chern-Simons density, these new effects were not observed. In the context of $3+1$ dimensions however, there do exist Chern-Simons densities,
namely the Higgs--Chern-Simons densities (also defined in \cite{Tchrakian:2015pka}),
whose dynamics does result in the non-standard relation of energy and global charges. This was demonstrated for the $SO(5)$ and $SU(3)$ Yang-Mills--Higgs monopoles in Refs.~\cite{Navarro-Lerida:2013pua,Navarro-Lerida:2014rwa}.
It is important in this connection, that Higgs--Chern-Simons dynamics cannot lead to the evolution of topological charge observed here for gauged Skyrmions influenced by Chern-Simons dynamics.

\bigskip
\bigskip
\noindent
   {\bf Acknowledgements}
We thank Valery Rubakov for valuable comments. We are grateful to Eugen Radu for substantial discussions and collaboration at the early stage of this work. We thank Yasha Shnir for bringing some interesting points to our attention, and pointing out some relevant references.

\appendix


\section{Topological charge of $SO(2)$ gauged $O(3)$ Skyrme system on $\R^2$}
\setcounter{equation}{0}
\renewcommand{\theequation}{A.\arabic{equation}}
The definition of the topological charge is independent of the dynamics of the gauge field. The definitions given
in this Appendix, apply to the following Appendices {\bf B} and {\bf C}, pertaining respectively to Maxwell
and Chern-Simons dynamics.

The topological charge for $SO(N)$ gauged $O(D+1)$ Skyrme system on $\R^D$, with $2\le N\le D$,
is given in \cite{Tchrakian:1997sj}
and is reviewed in Appendix {\bf B} of Ref,~\cite{Tchrakian:2015pka}. In the case $D=2$,
this definition coincides with that of
Ref.~\cite{Schroers:1995he}. Here, we revisit the $D=2$ case for completeness and, with more detail.



The $O(3)$ sigma model(s) are described by the scalar field $\f^a$, $a=\al,3$;
$\al=1,2$, which is subject to the constraint
\be
\label{constr}
|\f^a|^2=1\,.
\ee
From the outset we restrict our attention strictly to topologically stable Skyrmion solutions, which are characterised by the
asymptotics,
\be
\lim_{r \rightarrow 0} \phi^{3} =\mp 1\ ,\quad
\lim_{r \rightarrow\infty} \phi^{3} =\pm 1\,.\label{skasymp}
\ee
Moreover in what follows we will restrict to the upper signs in \re{skasymp}, for simplicity.

The prescription of gauging employed is
\bea
D_i\f^{\al}&=&\pa_i\f^{\al}+A_i(\vep\f)^{\al} \,,\label{covali}\\
D_i\f^3&=&\pa_i\f^3\label{cov3i}\,,
\eea
such that the component $\f^3$ is $gauge\ invariant$.

The winding number density prior to gauging is
\be
\label{3}
\varrho_0 =\vep_{ij} \vep^{abc}\pa_{i} \phi^{a} \: \pa_{j} \phi^{b}\phi^{c} \,,
\ee
which is $essentially\ total\ divergence$ but is $gauge\ variant$, while the density
\be
\label{4}
\varrho_G =\vep_{ij} \vep^{abc}D_{i} \phi^{a} \: D_{j} \phi^{b} \phi^{c} \,,
\ee
is $gauge\ invariant$ but is not a $total\ divergence$. Thus, neither qualifies as a topological charge density.

However, \re{3} and \re{4} are related through
\be
\label{5}
\varrho_G =\varrho_0 +2\vep_{ij} \pa_{i} (\phi^3A_{j})-\vep_{ij} \phi^3 F_{ij}\,.
\ee
Collecting the two $gauge\ invariant$ terms together, and the two $total\ divergence$ terms together, we propose
two equivalent definitions for a topological charge density
\bea
\tilde\varrho &=& \varrho_0+2\,\vep_{ij} \pa_{i} [(\phi^3)A_{j}]\label{ttp}\\
&=&\varrho_G+\vep_{ij} (\phi^3)F_{ij}\,.\label{gip}
\eea
The first line, \re{ttp}, is manifestly a total divergence while the second line, \re{gip} is manifestly gauge invariant.
Both are gauge invariant and total divergence, and are candidates for topological charge densities.

It is clear that the definitions \re{ttp}-\re{gip} can be extended by adding to (or subtracting from) each, the first
Chern-Pontryagin density~\footnote{This can be done in every even dimension $D=2n$ with the $n$-th Chern-Pontryagin density.}
$\vep_{ij}F_{ij}$, which is both total divergence and gauge invariant. Subtracting $\vep_{ij}F_{ij}=2\vep_{ij}\pa_iA_j$,
with a real constant coefficient $\upsilon$, results in the most general definition of the topological charge
\bea
\varrho &=& \varrho_0+2\vep_{ij} \pa_{i} [(\phi^3-\upsilon)A_{j}]\label{9a}\\
&=&\varrho_G+\vep_{ij} (\phi^3-\upsilon)F_{ij} \, . \label{9b}
\eea

Our considerations are restricted strictly to Skyrmion asymptotics \re{skasymp}, mostly to the upper signs there,
throughout.
For the purpose of establishing
Belavin inequalities, definition \re{9b} for the topological charge will be employed, while \re{9a}
will be employed to evaluate the topological charge itself.

\subsection{Topological charge of $SO(2)$ gauged Higgs system on $\R^2$}
Here, we return to the definition of the topological charge of the $SO(2)$ gauged Higgs scalar, in a manner analogous to
what was described above for the $O(3)$ Skyrme system.
The systematic way to achieve
this is to subject a Chern-Pontryagin (CP) density to dimensional descent to the desired dimension, in this case $\R^2$.
(See \cite{Tchrakian:2010ar} and references therein.)

This alternative approach was proposed in \cite{Tchrakian:2002ti} for the Higgs systems on $\R^2$ and $\R^3$.
Here we present only the case in $\R^2$.

The covariant derivative of the complex Higgs scalar $\vf=\f^1+i\f^2$, written for the real doublet $\f^a\ ,\ a=1,2$ is
\bea
D_i\f^{a}&=&\pa_i\f^{a}+A_i(\vep\f)^{a} \, . \label{cova}
\eea
In analogy with \re{3} and \re{4}, we define
the winding number density prior to gauging
\be
\label{3h}
\varrho_0 =\vep_{ij} \vep^{ab}\pa_{i} \phi^{a} \: \pa_{j} \phi^{b} \,,
\ee
which is $essentially\ total\ divergence$ but is $gauge\ variant$, and its covariantised version 
\be
\label{4h}
\varrho_G =\vep_{ij} \vep^{ab}D_{i} \phi^{a} \: D_{j} \phi^{b} \,,
\ee
which is $gauge\ invariant$ but is not a $total\ divergence$. Thus, neither qualifies as a topological charge density.

The relation between $\varrho_0$ and $\vr_G$, analogous with \re{5} is
\be
\label{5h}
\varrho_G =\varrho_0 -\vep_{ij} \pa_{i} (|\f^a|^2A_{j})+\frac12\vep_{ij}F_{ij}|\f^a|^2\,,
\ee
from which we deduce the two definitions of the topological charge density
\bea
\tilde\varrho &=& \varrho_0-\vep_{ij} \pa_{i} [|\f^a|^2A_{j}]\label{ttph}\\
&=&\varrho_G-\frac12\vep_{ij}|\f^a|^2 F_{ij}\,.\label{giph}
\eea
analogous with \re{ttp}-\re{giph}.

Finally, we add $\eta^2$ times the $1$st CP density $\frac12\vep_{ij}F_{ij}$ to state the final result
\bea
\varrho^{(\eta)} &=& \varrho_0+\vep_{ij} \pa_{i} [(\eta^2-|\f^a|^2)A_{j}]\label{9ah}\\
&=&\varrho_G+\frac12\vep_{ij}(\eta^2-|\f^a|^2) F_{ij}\,.\label{9bh}
\eea
Here, the real constant $\eta$ has dimension $[L^{-1}]$ like the Higgs scalar, unlike the dimensionless constant $v$ above,
\re{9a}-\re{9b}, for the Skyrme scalar.

\section{Topological lower bound for $SO(2)$ gauged Skyrmions in $2+1$ dimensions with
Maxwell dynamics}
\setcounter{equation}{0}
\renewcommand{\theequation}{B.\arabic{equation}}
The static Hamiltonian considered is the $SO(2)$ gauged planar Skyrme scalar with Maxwell dynamics,
\be
\label{HamMax}
{\cal H}_{(1)}
=\frac{1}{4}F_{ij}^2+\frac12\eta^2|D_i\f^a|^2+\eta^4V[\f^3] \,,
\ee
where $\eta$ be a constant~\footnote{$\eta$ is introduced for the purpose of keeping track of the dimensions of each term
in all the $O(3)$ sigma models considered,
and is set equal to a given value in the numerical calculations.} with dimension $L^{-1}$ and $V[\f^3]$ is the (dimensionless) potential,
which will be determined after requiring that \re{HamMax}
is bounded from below by either the topological charge density \re{gip} or \re{9b}.



The inequality giving rise to the $\varrho_G$ term in \re{gip} (or \re{9b}) is
\be
\label{ineq1}
\left|D_{i} \phi^a -\vep_{ij} \vep^{abc} D_{j} \phi^b \phi^c \right|^2\ge 0
\quad\Rightarrow\quad
|D_{i}\phi^a|^2 \ge\varrho_G\,.
\ee
and to reproduce the other term in \re{HamMax} we consider the inequality
\be
\label{ineq2}
(\vep_{ij}F_{ij}-\eta^2\,U)^2\ge0\quad\Rightarrow\quad  F_{ij}^2+\frac12\eta^4\,U^2\ge\eta^2\vep_{ij}F_{ij}U \,,
\ee
for some real gauge invariant function $U[\f^3]$ to be determined $via$ the inequalities \re{ineq1}-\re{ineq2},
which in turn imply the following Belavin inequality
\be
\label{Belavin}
\frac{1}{4}F_{ij}^2+\frac12\eta^2|D_i\f^a|^2+\frac18\eta^4U^2
\ge\frac12\eta^2\left[\vr_G+\frac12\vep_{ij}F_{ij}\,U\right]\,.
\ee
For the right hand side of \re{Belavin} to coincide with the topological charge density \re{9b}, fixes $U[\f^3]$ as
\be
\label{UMax}
U=2(\f^3-\upsilon)\,.
\ee
Finally, identifying the left hand side of \re{Belavin} with the static Hamiltonian \re{HamMax} fixes the potential
$V=\frac18U^2$,
\be
\label{Schpot}
V[\f^3]=\frac12(\f^3-\upsilon)^2\,.
\ee
So far the value of the constant $v$ is not fixed. 
    It is clear that with (the upper sign of) Skyrme asymptotics \re{skasymp}, the energy integral resulting from the left hand side of \re{Belavin}, will diverge except when one chooses $v=1$.

The topological charge is conveniently evaluated
using the expression \re{9a}, which in this case will get its value from the integral of the first term $\vr_0$ only,
since the second term vanishes at infinity by virtue of \re{skasymp} (upper sign). Thus the topological charge of
these vortices is equal to the ``baryon number'' of the planar Skyrmion in the gauge-decoupled limit.

The lower bound on the static Hamiltonian \re{HamMax} is now
\be
\label{fHamMax}
{\cal H}_{(1)}
=\frac{1}{4}F_{ij}^2+\frac12\eta^2|D_i\f^a|^2+\frac12\eta^4(\f^3-1)^2\ge\frac12\eta^2\,\vr\,,
\ee
which is saturated when both inequalities \re{ineq1} and \re{ineq2} are themselves saturated, resulting in the 
two first-order equations
\bea
D_{i} \phi^a &=&\vep_{ij} \vep^{abc} D_{j} \phi^b \phi^c \,, \label{bog1}\\
F_{ij}&=&\eta^2\,\vep_{ij}(\f^3-1)\,,\label{bog2}
\eea
which solve the second-order equations.

This lower bound is not invalidated if the potential \re{Schpot} is replaced by another potential which is
everywhere (within the acceptable range of values $(\f^3(0)=-1\ ,\f^3(\infty)=1)$) larger than \re{Schpot}. A natural example is
the ``pion mass'' potential $V=1-\f^3$.

More interesting however is the choice
\be
\label{JR}
V[\f^3]=\frac12\la(\f^3-1)^2\ ,\quad\la>0\,.
\ee
Denoting the resulting static Hamiltonian defined in terms of the potential \re{JR} as ${\cal H}_{(\la)}$, the lower bound
corresponding to \re{fHamMax} now reads
\bea
{\cal H}_{(\la)}&>&\frac12\eta^2\,\vr\ ,\quad\la>1\,,\label{JR1}\\
&>&\frac12\eta^2\,\la\,\vr\ ,\quad\la<1\,,\label{JR2}
\eea
following the analysis given in Ref.~\cite{Jacobs:1978ch} for the Abelian Higgs model.




\subsection{Topological lower bound for the Maxwell-Higgs case}
Our reason for considering the Higgs vortices analogous with the Skyrme vortices is, to emphasise the
marked difference between these. We have in mind here the usual Abelian Higgs model whose static
Hamiltonian is
\be
\label{HamAH}
{\cal H}=\frac14F_{ij}^2+\frac12|D_i\f^a|^2+\frac18(\eta^2-|\f^a|^2)^2 \,,
\ee
where the constant~\footnote{$\eta$ in Higgs models is not employed as an artifact like in the sigma models.}
$\eta$ is the VEV of the Higgs field.

The inequalities analogous with \re{ineq1}-\re{ineq2} in this case are
\bea
\left|D_{i} \phi^a -\vep_{ij} \vep^{ab} D_{j} \phi^b \right|^2\ge 0
\quad&\Rightarrow&\quad
|D_{i}\phi^a|^2 \ge\varrho_G \,, \label{ineq1h}\\
(\vep_{ij}F_{ij}-U)^2\ge0\quad&\Rightarrow&\quad  2F_{ij}^2+U^2\ge 2\vep_{ij}F_{ij}U\,.\label{ineq2h}
\eea
Combining \re{ineq1h} and \re{ineq2h}, and insisting that the resulting Belavin inequality be
\be
\label{Belineqh}
\frac14F_{ij}^2+\frac12|D_i\f^a|^2+\frac18(\eta^2-|\f^a|^2)^2\ge\frac12\vr^{(\eta)}\,,
\ee
$\vr^{(\eta)}$ given by \re{9bh}, results in
\bea
U&=&(\eta^2-|\f^a|^2)\,.\label{Uh}
\eea
It is clear that setting $\eta=0$, namely by opting for the definition \re{giph} for the topological charge,
the potential $V=U^2$ results in diverging energy
since the covariant constancy of the Higgs field implies that its magnitude vanishes at infinity.

\section{Topological lower bound for $SO(2)$ gauged Skyrmions in $2+1$ dimensions with
Chern-Simons dynamics}
\setcounter{equation}{0}
\renewcommand{\theequation}{C.\arabic{equation}}
The topological lower bound on the energy of the static solutions to the Lagrangian
\be
\label{CSLag}
{\cal L}_{(1)}=\ka\vep^{\la\mu\nu}A_\la F_{\mu\nu}+\frac12\eta^2|D_\mu\f^a|^2-\eta^4\,V[\f]\ ,\quad \mu=0,i \;\quad i=1,2\,,
\ee
are sought. Here, $\ka$ is the coupling strength of the Chern-Simons (CS) term and like $\eta$ has dimension $[L^{-1}]$.
The gauging prescription is that of the Maxwell case, \re{coval}-\re{cov3}, such that the static Hamiltonian $2T_{00}$
of \re{CSLag} is expressed as
\be
\label{CSHam}
{\cal H}=\frac12\eta^2\left(|D_i\f^a|^2+A_0^2|\f^\al|^2\right)+\eta^4\,V[\f]\ ,\quad a=\al,3\ ;\quad\al=1,2\,.
\ee
The Gauss Law equation ($0$-component of the Maxwell equation) is
\be
\label{Gauss}
A_0=-\frac{2\ka}{\eta^2|\f^\al|^2}\vep_{ij}F_{ij\,.}
\ee
Substituting \re{Gauss} in \re{CSHam} we have the final form of the static Hamiltonian density
\be
\label{CSHamfin}
{\cal H}_{(1)}=4\left(\frac{\ka}{\eta}\right)^2\left(\frac{F_{ij}^2}{|\f^\al|^2}\right)
+\frac12\eta^2\,|D_i\f^a|^2+\eta^4\,V[\f]\ ,\quad a=\al,3\ ;\quad\al=1,2\,.
\ee
So far, the potential $V[\f]$ is not specified.
This can be done once the Belavin inequalities that give the
lower bound on the energy, namely on the integral of \re{CSHamfin}, are stated. This depends on the definition of the
topological charge density.

The inequality giving rise to the $\varrho_G$ term in \re{gip} (or \re{9b}) is of course \re{ineq1}, again.
To account for the term $F_{ij}^2/|\f^\al|^2$ term in \re{CSHamfin}, consider the inequality
\bea
\left(\frac{\ka}{\eta|\f^\al|}\vep_{ij}F_{ij}-\eta^2|\f^\al|U\right)^2\ge 0
&\Rightarrow&2\left(\frac{\ka}{\eta}\right)^2\left(\frac{F_{ij}^2}{|\f^\al|^2}\right)
+\eta^4|\f^\al|^2U^2\ge 2\ka\eta\vep_{ij}F_{ij}\,U \, .
\label{Bineq2}
\eea
Combining \re{ineq1} and \re{Bineq2} we have the Belavin inequality
\be
\label{Bineq}
4\left(\frac{\ka}{\eta}\right)^2\frac{F_{ij}^2}{|\f^\al|^2}+\frac12\eta^2|D_i\f^a|^2+
2\eta^4|\f^\al|^2U^2\ge\frac12\eta^2\left[\vr_G+8\left(\frac{\ka}{\eta}\right)\vep_{ij}F_{ij}\,U\right]\,.
\ee
Requiring that the expression in the square brackets on the right hand side of \re{Bineq} be the topological charge $\vr$,
given by \re{9b}, fixes $U$ as
\be
\label{UCS}
U=\frac18\left(\frac{\eta}{\ka}\right)(\f^3-\upsilon)\,,
\ee
and identifying the left hand side of \re{9b} with the the static Hamiltonian density \re{CSHamfin} results in the following
potential $V=2|\f^\al|^2U^2$,
\be
\label{VCS}
V[\f]=\frac{1}{32}\left(\frac{\eta}{\ka}\right)^2|\f^\al|^2(\f^3-\upsilon)^2\,.
\ee

So far the value of the constant $v$ in \re{VCS} is not fixed. In contrast with the Maxwell dynamics above, here
the energy integral will not diverge if the choice $v=1$ is not made. This is seen immediately by inspecting \re{VCS},
which must decay at infinity since
\[
\lim_{r\to\infty}|\f^\al|^2=0\,,
\]
for any value of $v$. The question as to
what interesting choices for $v$ can be made depends on the detailed analysis of the solutions.

What is important here is that when calculating the topological charge by evaluating the integral of \re{9a}, the result will not
be limited to the  winding number, namely the integral of $\vr_0$,
except when the choice $v=1$ is made. This is in contrast to the case of Maxwell dynamics considered above.

The final result is the Belavin inequality
\be
\label{Bfinineq}
{\cal H}_{(1)}=4\left(\frac{\ka}{\eta}\right)^2\frac{F_{ij}^2}{|\f^\al|^2}+\frac12\eta^2|D_i\f^a|^2+
\frac{1}{32}\left(\frac{\eta^3}{\ka}\right)^2|\f^\al|^2(\f^3-\upsilon)^2\ge\frac12\eta^2\,\vr\,,
\ee
which is saturated if the two inequalities \re{ineq1} and \re{Bineq2} are saturated, $i.e.$, when the first-order equations
\bea
\vep_{ij}D_{i} \f^a &=&\vep^{abc} D_{j} \phi^b \phi^c \,,  \label{bogCS1}\\
8\ka^2\vep_{ij}F_{ij}&=&\eta^4\,|\f^\al|^2(\f^3-\upsilon)\,,\label{bogCS2}
\eea
are satisfied, which solve the second-order equations.

Again, as in the Maxwell case above, the lower bound \re{Bfinineq} is not invalidated if the potential \re{VCS} is replaced by
\be
\label{JRCS}
V[\f]=\frac{1}{32}\,\la\,\left(\frac{\eta}{\ka}\right)^2|\f^\al|^2(\f^3-\upsilon)^2\ ,\quad\la>0\,.
\ee
Denoting the static Hamiltonian defined in terms of the potential \re{JRCS} as ${\cal H}_{(\la)}$, the lower bound
corresponding to \re{Bfinineq} now reads
\bea
{\cal H}_{(\la)}&>&\frac12\eta^2\,\vr\ ,\quad\la>1\,,\nonumber\\
&>&\frac12\eta^2\,\la\,\vr\ ,\quad\la<1\,,\nonumber
\eea
formally the same as \re{JR1}-\re{JR2}.

\subsection{Topological lower bound for the Chern-Simons--Higgs case}
Again, we consider the Higgs analogue of the Chern-Simons--Skyrme vortices discussed above, to
emphasise the essential difference between them.

Starting with the Lagrangian
\be
\label{CSLagh}
{\cal L}=\ka\vep^{\la\mu\nu}A_\la F_{\mu\nu}+\frac12\eta^2|D_\mu\f^a|^2-V[|\f^a|^2]\ ,
\quad \mu=0,i \;\quad i=1,2\ , \ a=1,2\,,
\ee
one finds from the Gauss Law equation that
\be
\label{Gaussh}
A_0=-\frac{2\ka}{|\f^a|^2}\vep_{ij}F_{ij\,.}
\ee
whence the static Hamiltonian can be expressed as
\be
\label{CSHamfinh}
{\cal H}=4\ka^2\left(\frac{F_{ij}^2}{|\f^a|^2}\right)
+\frac12|D_i\f^a|^2+V\,.
\ee
Here, in addition to the inequality \re{ineq1h}, one employs
\bea
\left(\frac{\ka}{|\f^a|}\vep_{ij}F_{ij}-|\f^a|U\right)^2\ge 0
&\Rightarrow&2\ka^2\left(\frac{F_{ij}^2}{|\f^a|^2}\right)+|\f^a|^2U^2\ge 2\ka\vep_{ij}F_{ij}\,U\,.
\label{Bineq2h}
\eea
Combining \re{ineq1h} and \re{Bineq2h} one ends up with the Belavin inequality
\be
\label{Bineqh}
4\ka^2\frac{F_{ij}^2}{|\f^a|^2}+\frac12|D_i\f^a|^2+
2|\f^a|^2U^2\ge\frac12\eta^2\left[\vr_G+8\left(\frac{\ka}{\eta}\right)\vep_{ij}F_{ij}\,U\right]\,.
\ee
and insisting that the right hand side of \re{Bineqh} be equal to the topological charge density \re{9b}, fixes $U$ as
\be
\label{UCSh}
U=\frac{1}{16\ka}(\eta^2-|\f^a|^2)\,.
\ee
It follows that the potential $V$ is now
\be
\label{VCSh}
V[|\f^a|^2]=\frac{2}{16^2\ka^2}|\f^a|^2(\eta^2-|\f^a|^2)\,.
\ee
This is known from the work of \cite{Hong:1990yh,Jackiw:1990aw}.

What must be emphasised here is that setting $\eta=0$
results in divergent energy since at infinity $|\f^a|^2\to\eta^2$, in contrast with the Skyrme case above where
$|\f^\al|^2\to 0$, which allows the choice $v=0$ there.

\begin{small}

\end{small}

\end{document}